\documentclass[10pt,journal,compsoc]{IEEEtran}

\usepackage{amsmath,amssymb,amsfonts}
\usepackage{algorithmic}
\usepackage{graphicx}
\usepackage{textcomp}
\usepackage{multirow}
\usepackage{xcolor}
\usepackage{bm}
\usepackage{makecell}
\usepackage{booktabs}
\usepackage{amsmath}
\usepackage{xcolor}
\usepackage[linesnumbered,ruled,vlined]{algorithm2e}

\ifCLASSOPTIONcompsoc
  \usepackage[nocompress]{cite}
\else
  \usepackage{cite}
\fi

\ifCLASSINFOpdf
\else
\fi
\usepackage{url}

\begin{document}

\title{StagedVulBERT: Multi-Granular Vulnerability Detection with a Novel Pre-trained Code Model}

\author{
	Yuan Jiang, Yujian Zhang,
    Xiaohong Su, Christoph Treude and
    Tiantian Wang
\IEEEcompsocitemizethanks{
\IEEEcompsocthanksitem Y. Jiang (jiangyuan@hit.edu.cn), Y. Zhang (23S003081@stu.hit.edu.cn), X. Su (sxh@hit.edu.cn), and T. Wang (wangtiantian@hit.edu.cn), are with the School of Computer Science and Technology, Harbin Institute of Technology, Harbin,
Heilongjiang, 15001.\protect\\

\IEEEcompsocthanksitem C. Treude is with the School of Computing and Information Systems, Singapore Management University,
Singapore.\protect\\
E-mail: ctreude@smu.edu.sg
}

}

\IEEEtitleabstractindextext{
\begin{abstract}

The emergence of pre-trained model-based vulnerability detection methods has significantly advanced the field of automated vulnerability detection. However, these methods still face several challenges, such as difficulty in learning effective feature representations of statements for fine-grained predictions and struggling to process overly long code sequences. To address these issues, this study introduces StagedVulBERT, a novel vulnerability detection framework that leverages a pre-trained code language model and employs a coarse-to-fine strategy. The key innovation and contribution of our research lies in the development of the CodeBERT-HLS component within our framework, specialized in hierarchical, layered, and semantic encoding. This component is designed to capture semantics at both the token and statement levels simultaneously, which is crucial for achieving more accurate multi-granular vulnerability detection. Additionally, CodeBERT-HLS efficiently processes longer code token sequences, making it more suited to real-world vulnerability detection. Comprehensive experiments demonstrate that our method enhances the performance of vulnerability detection at both coarse- and fine-grained levels. Specifically, in coarse-grained vulnerability detection, StagedVulBERT achieves an F1 score of 92.26\%, marking a 6.58\% improvement over the best-performing methods. At the fine-grained level, our method achieves a Top-5\% accuracy of 65.69\%, which outperforms the state-of-the-art methods by up to 75.17\%.


\end{abstract}

}

\maketitle

\IEEEdisplaynontitleabstractindextext

%
\IEEEpeerreviewmaketitle

\IEEEraisesectionheading{\section{Introduction}\label{sec:introduction}}

\IEEEPARstart{S}{oftware} vulnerability discovery is a classic problem that becomes challenging due to its undecidable nature in the general case~\cite{Yamaguchi2014,rice1953classes}.
Recently, with the further development of deep learning (DL) technologies, data-driven vulnerability detection with automatic extraction of vulnerability patterns has attracted many researchers to invest in this prominent research field.
These approaches~\cite{Russell} are often formulated as a classification task, which takes code features as input, mapping them to the binarization of the probability distribution that indicates whether or not the program contains a vulnerability. Although these methods (e.g., Devign~\cite{zhou2019devign}, Reveal~\cite{chakraborty2021deep}, IVDetect~\cite{li2021vulnerability}, AMPLE~\cite{wen2023vulnerability}, VulDeePecker~\cite{li2016vulpecker} and SySeVR~\cite{li2018sysevr}) have improved vulnerability detection by considering complex structural information and various code representations, their effectiveness remains constrained. The primary reason is that models trained solely on limited vulnerability datasets face difficulties in extracting complex semantic patterns, leading to inadequate performance and restricted practicality in detecting software vulnerabilities in real-world scenarios.

To address such shortcomings, the state-of-the-art methods LineVul~\cite{fu2022linevul}, VulBERTa~\cite{hanif2022vulBERTa} and SVulD~\cite{ni2023distinguishing} adopt a new learning paradigm of ``pre-training and fine-tuning" with the Pre-trained Code Language (PCL) models for source code representation and vulnerability detection. Concretely, these approachess, different from other DL-based detection methods, involve self-supervised pre-training on large-scale open-source code datasets, followed by fine-tuning on vulnerability datasets, thereby achieving superior performance compared to traditional models. A recent empirical study also demonstrates that the transformer-based method LineVul achieves the best performance among the 11 cutting-edge DL-based detection methods~\cite{steenhoek2023empirical}. 
In addition, among transformer-based methods, LineVul is the only one designed for multi-granular vulnerability detection, while the others (e.g., SVulD and VulBERTa) are limited to coarse-grained vulnerability detection.

Although current efforts made by pre-trained model-based approaches (e.g., LineVul) have improved the detection capabilities, they still exhibit the following weaknesses:
(1) they mainly treat code as a sequence of tokens and adopt the basic transformer architecture to capture the relationships between tokens. However, relying solely on token-level features makes it difficult to achieve excellent performance in statement-level vulnerability detection, as the cases provided in Section 2.1 show;
(2) they, if designed for multi-granular vulnerability,  solely employ function-level (coarse-grained) labels for training models to achieve fine-grained detection in unseen programs. The objective gaps between training and prediction~\cite{cui2021directqe} may lead to poor performance in pinpointing vulnerable lines;
(3) they predominantly build on the existing PCL models (e.g., CodeBERT~\cite{feng2020codeBERT,fu2022linevul} or UniXcoder~\cite{ni2023distinguishing}), which are limited by sequence lengths -- typically a maximum of 512 tokens. Thus, code functions that exceed the maximum length (i.e., 512) will simply be truncated, which may result in the loss of useful information.

To overcome the limitations of the state-of-the-art method, in this study, we propose \textit{StagedVulBERT}, \textit{a staged vulnerability detection framework employing CodeBERT-HLS}, which offers several key advantages.
Firstly, it leverages the proposed CodeBERT-HLS to integrate low-level (i.e., token-level) features and high-level (i.e., statement-level) semantic features to accurately represent programs.
Secondly, it directly uses coarse- and fine-grained labels to construct distinct binary classifiers through supervised learning for coarse- and fine-grained vulnerability detection.
Thirdly, it is equipped with a novel algorithm that can segment the token sequence and then effectively apply various components of CodeBERT-HLS to process longer token sequences.
The three innovations respectively address the aforementioned limitations of previous transformer-based works.
In StagedVulBERT, CodeBERT-HLS serves as the core component, representing an advanced pre-trained model specialized in \textbf{H}ierarchical \textbf{L}ayered \textbf{S}emantic encoding for source code. A key innovation in CodeBERT-HLS is that it employs an architecture featuring segmented Transformer encoders. This design is crucial for simultaneously capturing semantics at both token and statement levels and for processing longer token sequences efficiently. 
Additionally, a novel pre-training strategy, the Masked Statement Prediction (MSP) model, has been developed for CodeBERT-HLS to enhance pre-training effectiveness. 
Consequently, on real-world vulnerability datasets, our approach has been proven to more accurately detect vulnerabilities compared to existing methods through extensive evaluations.

The main \textbf{contributions} of our paper are:
\begin{itemize}

\item \textbf{New vulnerability detection framework}. We construct a novel vulnerability detection framework, StagedVulBERT, which not only processes longer token sequences than existing models but also performs the vulnerability detection task with high performance at both the coarse- and fine-grained levels.

\item \textbf{New pre-trained code language model for vulnerability detection}. We introduce a new pre-trained code language model CodeBERT-HLS specifically designed for multi-granular vulnerability detection, which has the advantage of modeling code semantics at different levels of granularity. 

\item \textbf{New pre-training task for code representation model}. We develop a novel pre-training task tailored to enhance the model's comprehension of code syntax and semantic relationships, which is crucial for achieving high accuracy in the downstream vulnerability detection task.

\end{itemize}


\section{Background and Motivation}
\label{background}
\subsection{Motivating Example}
\begin{figure}[htbp]
	\centerline{
		\includegraphics[width=0.5\textwidth]{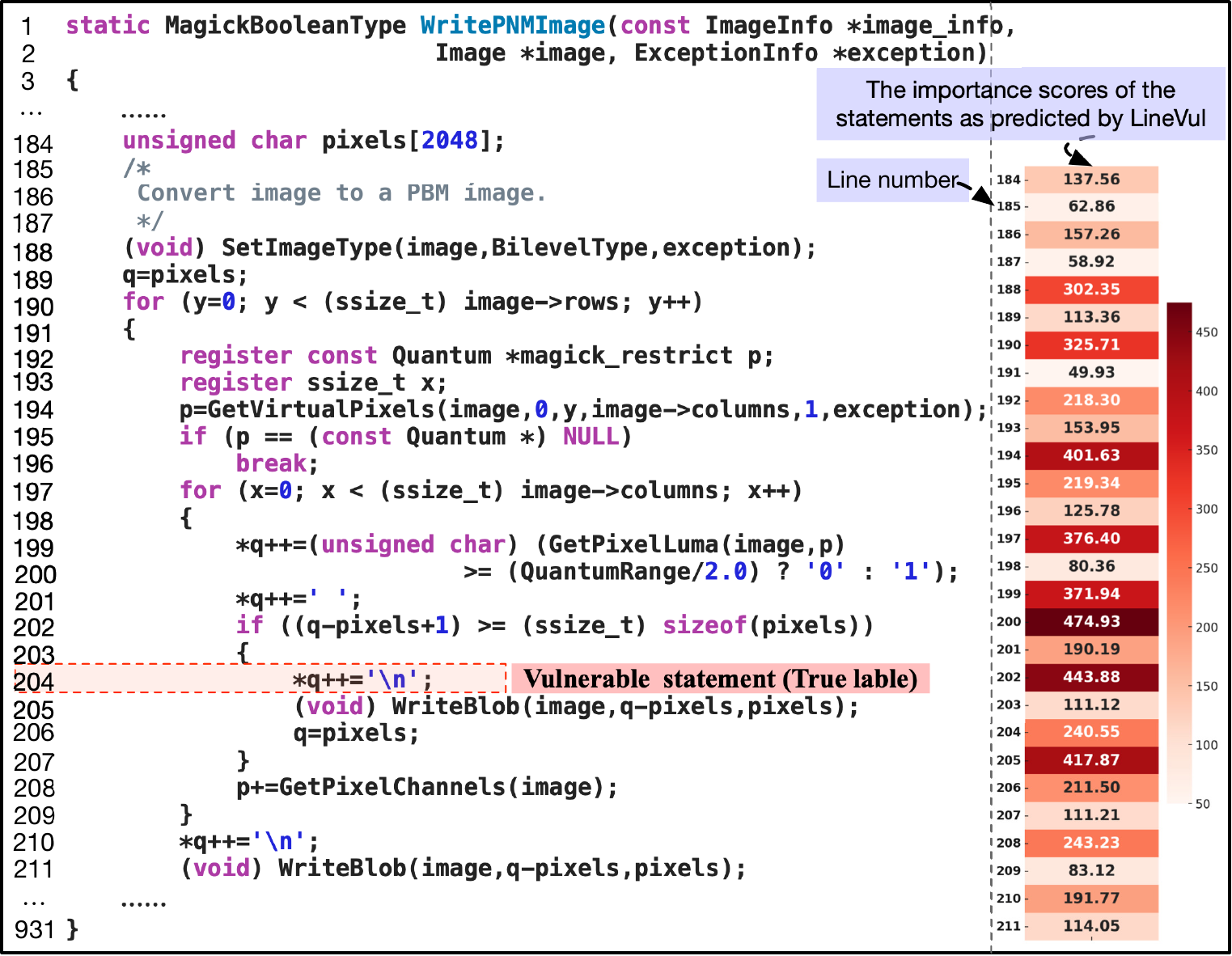}}
	\caption{A real-world example (i.e., CVE-2019-13304) of stack-based buffer overflow, which occurs from writing excess data to an array due to improper buffer size check.}
	\label{VulExample}
\end{figure}
Fig.~\ref{VulExample} depicts a vulnerability (CVE-2019-13304) in ImageMagick, which is classified as CWE-787 in the National Vulnerability Database. This example shows a stack-based buffer overflow due to improper buffer size checks in lines 197-207. The code declares an array for image pixels (line 184) and erroneously allows overflow when writing newline characters (line 204, marked in red dashed boxes), leading to memory corruption. 
When applying the state-of-the-art method LineVul for vulnerability localization on specific segments of the program (lines 184-211), the results are displayed as a heatmap, which is illustrated on the right side of Fig.~1. In this heatmap, each element represents the importance score assigned by LineVul to each line of code, indicating its contribution to the vulnerability detection result.

From the example, we can make three observations. The first is that vulnerabilities in real-world projects (e.g., Fig.~\ref{VulExample}) are often complex to understand and may require considerable expertise for detection. However, existing DL-based methods were mainly trained on vulnerability datasets (e.g., big\_vul with approximately 180,000 samples), which are relatively small compared to other available general code databases (e.g., CodeSearchNet with approximately 8,000,000 samples) due to the high cost of labeling. Recent works~\cite{fu2022linevul,hanif2022vulBERTa} have demonstrated that first pre-training code language models on large-scale code datasets like CodeSearchNet can improve the semantic understanding of vulnerable code, and then fine-tuning these models on vulnerability datasets can enhance their detection capabilities. However, most of these methods are based on the original transformer architecture, which does not perform well for multi-granular vulnerability detection. Therefore, in this study, we develop a new PCL model, CodeBERT-HLS, tailored for multi-granular vulnerability detection, and design suitable pre-training tasks.

The second observation is that some vulnerable programs can be very long, spanning thousands of tokens. Therefore, most DL-based methods struggle to detect such vulnerabilities due to a limited input token number. For example, the best-performing transformer-based method, LineVul, for vulnerability detection only accepts up to 512 tokens~\cite{fu2022linevul}, which cannot detect the vulnerable example in Fig.~\ref{VulExample} because the vulnerable statement exceeds the first 512 tokens.
Although recently proposed Large Language Models (LLMs) seem to address the challenge of limited length, they perform worse than transformer-based methods~\cite{yin2024multitask,fu2023chatgpt}. Therefore, in this study, we develop an advanced algorithm tailored to the proposed transformer-based method for effective, long-sequence vulnerability detection.

The third observation is that most transformer-based methods for vulnerability detection perform poorly when applied to the fine-grained level.
As the heatmap in Fig.~\ref{VulExample} demonstrates, in the code segment between lines 184-211, the most likely vulnerability predicted by LineVul is at Line 200, rather than the actual vulnerable statement at Line 204. 
This indicates that the state-of-the-art method's fine-grained vulnerability prediction is not sufficiently accurate.
The reasons for the failed detection are: (1) LineVul trains CodeBERT only at a coarse level, causing misalignment with fine-grained detection goals~\cite{cui2021directqe}; (2) LineVul generates features at the token and program levels, which are not tailored for statement-level vulnerability detection.
Fine-grained labeling marks specific vulnerable lines within functions, and all statements corresponding to fine-grained labels form the vulnerability-relevant structure of the code snippet.
Therefore, employing fine-grained labels in model training can reduce the above training-predicting gap and provide more precise signs to guide the model in learning about the specific code statements causing vulnerabilities. To achieve this, we propose CodeBERT-HLS to generate statement representations, which, along with fine-grained labels, help construct a more accurate fine-grained detection model.

We emphasize once more that although recent LLMs have addressed many challenges of DL- or transformer-based methods mentioned in the above three observations, empirical studies~\cite{yin2024multitask,fu2023chatgpt} suggest they do not outperform specific transformer-based methods like LineVul in vulnerability detection, even post-fine-tuning. Hence, given that current LLMs are suboptimal in detecting vulnerabilities, improving transformer-based methods to address these challenges is a valuable research direction.

\subsection{Background}
\label{bg-rl}
Since this study aims to enhance transformer-based methods, a category of DL-based methods, we begin with an overview of DL-based methods before delving into the specifics of transformers.
\subsubsection{DL-based Vulnerability Detection Methods}
Due to programs typically being written by humans, they share many important properties with natural language text (e.g., repetitive, predictable, and long-term dependencies)~\cite{hindle2012naturalness,dam2018automatic}. Thus, more and more researchers focus on capturing program features by means of statistical language models.
Numerous studies also demonstrate that neural networks achieve superior results with source code, similar to their performance with natural language~\cite{dam2016deep,hellendoorn2017deep}. 
Moreover, a DL model, such as LSTM, itself is a good feature extractor, hence, it can extract features in an automatic way instead of designing various sophisticated feature extraction rules~\cite{lee2017applying}. However, most DL-based models struggle to effectively learn vulnerability patterns for both coarse- and fine-grained detection. Therefore, further research into advanced network architectures suited for multi-granular vulnerability detection is necessary. To make the concept of ``multi-granular vulnerability detection" more clear, we give its definition as follows: Multi-granular vulnerability detection involves identifying and classifying vulnerabilities at various levels of code abstraction, such as function and statement levels, using techniques like static analysis, dynamic analysis, or data-driven methods.

\subsubsection{Transformer-based Vulnerability Detection Methods}
Inspired by the significant success of pre-training in NLP, researchers have begun applying transformers to vulnerability detection with extensive empirical evidence demonstrating their effectiveness over traditional DL-based methods~\cite{steenhoek2023empirical}. In this study, following previous work~\cite{yin2024multitask}, we categorize transformer-based methods used for vulnerability detection into two groups: LLM-based and specific transformer-based. LLM-based methods, well-trained with billions of parameters and training samples, generally perform well in various software engineering tasks that require high language comprehension and generation capabilities. 
The LLMs recently proposed and utilized extensively include DeepSeek-Coder~\cite{guo2024deepseek}, CodeLlama~\cite{roziere2023code}, StarCoder~\cite{li2023starcoder}, WizardCoder~\cite{luo2023wizardcoder}, Mistral~\cite{jiang2023mistral}, Phi-2~\cite{javaheripi2023phi}, and ChatGPT~\cite{ouyang2022training}.

Specific transformer-based approaches have emerged to utilize PCL models for detecting vulnerabilities. These methods also adhere to strategies that first involve pre-training, followed by fine-tuning. 
Notable PCL-based vulnerability detection methods include LineVul, VulBERTa and SVulD. Most of these methods are based on the basic transformer architecture and well-known pre-training tasks (e.g., Masked Language Modeling (MLM)~\cite{2018bert}, ULM~\cite{brown2020language}, and Contrastive Learning (CL)~\cite{jain2020contrastive}. Note that pre-training tasks refer to specific training objectives designed to help the model learn general representations from large, unlabelled datasets via self-supervised learning.

In this study, we focus on enhancing specific transformer-based methods, which empirical evidence indicates as superior to other LLM-based approaches~\cite{yin2024multitask,fu2023chatgpt}, yet they exhibit the following limitations:
(1) most PCL models in transformer-based methods target coarse-grained vulnerability detection, with fewer addressing detection at multiple granularities; (2) existing PCL models typically generate feature representations at either the token or program level, which challenges statement-level vulnerability detection due to the lack of specific statement features; and
(3) most pre-training tasks are designed for the standard transformer architecture and may not be suitable when this architecture is modified.
Therefore, this paper develops a novel PCL model and an associated pre-training task, aiming to enhance multi-granular vulnerability detection to address these limitations.

\section{The overall framework of our method}
\label{overall}
\begin{figure*}[htbp]
	\centerline{
		\includegraphics[width=1.0\textwidth]{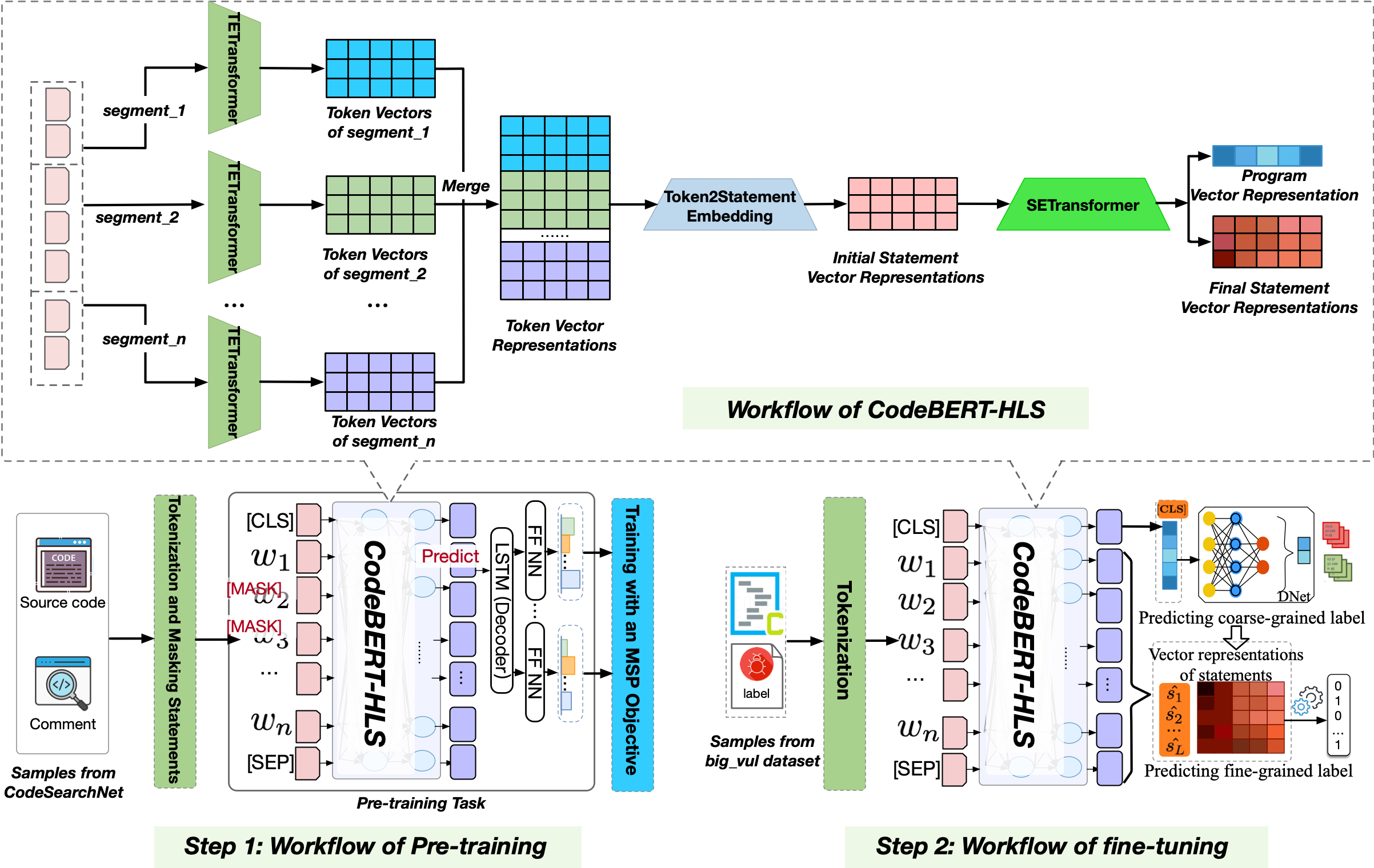}}
	\caption{Workflow of the proposed StagedVulBERT, which includes two steps:  pre-training and fine-tuning. The Workflow of CodeBERT-HLS is illustrated at the top of the figure.}
	\label{workflow_comp}
\end{figure*}

To address the challenge of accurately extracting vulnerability semantics from small datasets, we adopt a strategy of first pre-training on large-scale open-source code datasets, followed by fine-tuning on a relatively small vulnerability dataset, a method previously proven to enhance detection accuracy effectively~\cite{fu2022linevul}. Our approach, distinct from existing pre-training-based methods~\cite{fu2022linevul,hanif2022vulBERTa}, introduces a novel transformer-based model architecture CodeBERT-HLS and a pre-training task MSP, which are particularly suitable for multi-granular vulnerability detection. 
The work process of our model architecture, CodeBERT-HLS, is illustrated at the top of Fig.~\ref{workflow_comp}, while the workflow of pre-training and fine-tuning CodeBERT-HLS is shown at the bottom of Fig.~\ref{workflow_comp}.

\subsection{Workflow of CodeBERT-HLS} 
\begin{figure*}[htbp]
	\centerline{
		\includegraphics[width=1.0\textwidth]{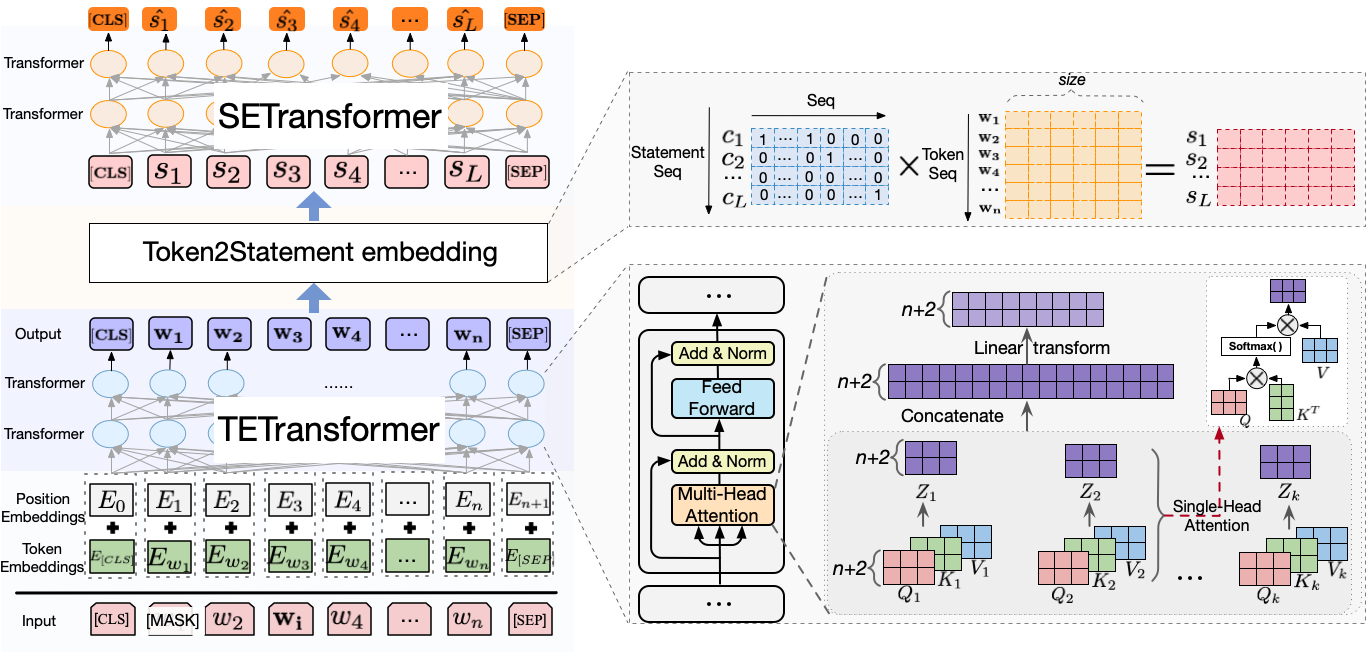}}
	\caption{The overall framework of the CodeBERT-HLS method, which is composed of the following three components: TETransformer, Token2Statement embedding, and SETransformer.}
	\label{CodeBERT-HLS}
\end{figure*}

Unlike most related work~\cite{fu2022linevul,hanif2022vulBERTa} that uses existing PCL models (e.g., CodeBERT) capturing only token-level information from token sequences, our approach involves exploring staged code representation with the proposed PCL model CodeBERT-HLS to capture information at both the token-level and statement-level, thereby facilitating a deeper understanding of vulnerable programs.
The architecture of CodeBERT-HLS is depicted at the top of Fig.~\ref{CodeBERT-HLS}, and its workflow is as follows: First, the code snippet is treated as a sequence of code tokens, which are then split into $n$ segments. The Token Encode Transformer (TETransformer) is used to learn features for each segment, generating token vector representations. Next, all token vectors from different segments are merged into a matrix. This matrix is then passed as input to the Token2Statement embedding layer to obtain the initial vector representation of each statement. Finally, CodeBERT-HLS leverages the Statement Encode Transformer (SETransformer) to capture dependency relationships among statements, which allows for the generation of more accurate statement vectors. The reason for splitting token sequences into multiple segments is to effectively handle long code token sequences. More details can be found in Section~\ref{sec:algorithm}.

\subsection{Workflow of Pre-training and Fine-tuning}
A pre-training process is employed in this study to learn a general-purpose code representation. However, existing pre-training tasks, such as MLM, are only suitable for training models that generate token-level representations. This makes them unsuitable for training CodeBERT-HLS, which is designed to output statement-level representations. Therefore, we design a new pre-training task MSP to align with the architecture of CodeBERT-HLS.
The workflow involves three steps: (1) masking tokens from the selected statements in a code sequence and inputting the masked sequence into CodeBERT-HLS, (2) generating a context-based representation for each masked statement, and (3) predicting the tokens of the masked statements based on the statement's representation and calculating the loss to adjust the network parameters.

After pre-training, we fine-tune the pre-trained CodeBERT-HLS model on the big\_vul dataset to get suitable weights for our vulnerability detection task. The workflow can be described as follows:
(1) inputting a code snippet into CodeBERT-HLS to obtain representations for each statement and the entire code snippet, which can be regarded as feature vectors for vulnerability detection.
(2) employing an FFNN named DNet that takes the feature vector of a code snippet as input and predicts the corresponding vulnerability label at the coarse-grained level.
(3) once a code snippet is predicted as vulnerable, constructing a fine-grained network that takes a statement vector produced by the SETransformer as input and outputs a probability distribution over the label space $\mathcal{L}$.
(4) calculating the cross-entropy losses based on the coarse and fine prediction results to fine-tune the network parameters.

\subsection{Advantages Over Other Methods} 
Compared to state-of-the-art methods based on PCL models (e.g., LineVul), our proposed approach offers several distinct advantages. Firstly, our method is capable of generating feature representations at both the program and statement levels, making it suitable for downstream multi-granular vulnerability detection. Secondly, it employs fine-grained labels to construct a binary classifier based on statement feature representations. This enables the automatic and more accurate determination of each statement's relevance to vulnerabilities. Thirdly, our method is capable of efficiently processing long token sequences, which enables it to extract more semantic information from the program and achieve better generalization in real-world scenarios. Fourthly, we introduce a pre-training task that is more suitable for our model architecture than the existing tasks (e.g., MLM) are, to enhance our model's understanding of general code semantics through self-supervised learning.

\section{Our approach}
\label{approach}
This section focuses on introducing the core components of CodeBERT-HLS and the pre-training task tailored specifically for CodeBERT-HLS. Additionally, this section explains how CodeBERT-HLS handles long code token sequences.

\subsection{Token Encode Transformer (TETransformer)}
\label{SENet}
TETransformer is designed to capture the semantic relationships between tokens in a program, generating a vector representation for each token in a code fragment. The TETransformer adopts an architecture identical to that of BERT~\cite{2018bert} and RoBERTa~\cite{liu2019roberta}, employing a multi-headed self-attention mechanism. This mechanism allows for simultaneous processing of multiple attention heads, each focusing on different aspects of the input token relationships. 
As a result, it enables the model to capture diverse contexts and relationships within the input token sequence. This enhanced capability allows for extracting more accurate semantics of vulnerable programs, significantly improving the model's overall performance in detecting vulnerabilities. A formal description of the TETransformer model is as follows:

\begin{equation}\label{eq2}
	G^{n}=LN\left(\text {MultiAttn }\left(H^{n-1}\right)+H^{n-1}\right) \\
\end{equation}
\begin{equation}\label{eq2}
	H^{n}=L N\left(F F N\left(G^{n}\right)+G^{n}\right)
\end{equation}

where \textit{MultiAttn} denotes the multi-headed self-attention mechanism, \textit{FFN} represents a two-layer FFNN, and \textit{LN} signifies layer normalization operations. For the $ n^{th} $ Transformer layer, the output of the multi-headed self-attention, $ \hat{G}^{n} $, is computed as follows:

\begin{equation}\label{eq2}
	Q_{i}=H^{n-1} W_{i}^{Q}, K_{i}=H^{n-1} W_{i}^{K}, V_{i}=H^{n-1} W_{i}^{V}
\end{equation}
\begin{equation}\label{eq2}
	head_{i}=\operatorname{softmax}\left(\frac{\mathrm{Q}_{\mathrm{i}} \mathrm{K}_{\mathrm{i}}^{\mathrm{T}}}{\sqrt{\mathrm{d}_{\mathrm{k}}}}\right) \mathrm{V}_{\mathrm{i}}
\end{equation}
\begin{equation}\label{eq2}
	\hat{G}^{n}=\left[\text {head}_{1} ; \ldots ; \text {head}_{u}\right] W_{n}^{O}
\end{equation}

where the output of the previous layer, \( H^{n-1} \in \mathbb{R}^{|X| \times d_{h}} \), is linearly projected using learnable model parameter matrices \( W_{i}^{Q}, W_{i}^{K}, W_{i}^{V} \in \mathbb{R}^{d_{h} \times d_{k}} \) into a triplet of queries \( Q \), keys \( K \), and values \( V \). Here, \( u \) denotes the number of heads in the multi-headed mechanism, \( d_k \) represents the dimensionality of the feature vector output by each head, and \( W_{n}^{O} \in \mathbb{R}^{ud_{k} \times d_{h}} \) is a learnable model parameter. The parameter \( W_{n}^{O} \) is used to linearly map the concatenated output of the multi-heads into an output vector with the same dimensionality as that of the input to the Transformer layer.

\subsection{Token2Statement Embedding}
The obtained sequence of token vectors is fed into the Token2Statement layer to calculate the initial representation for each statement. In this section, we provide the following three strategies used for Token2Statement Embedding, with the most effective strategy to be identified via experimental evaluation.

\begin{figure}
	\centering{\includegraphics[width=0.5\textwidth]{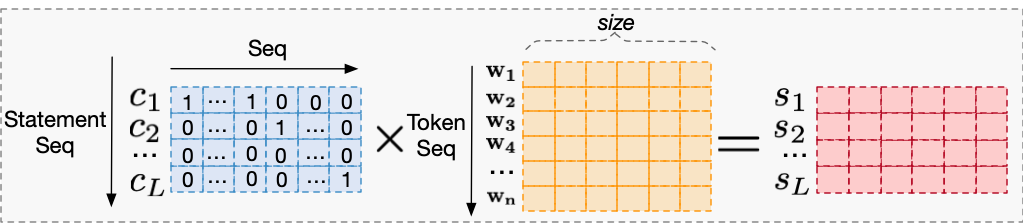}}
	\caption{Initialization of statement vector representations based on token vectors and statement-token correspondence}
	\label{fig:Token2Statement1}
\end{figure}

(1) The first strategy, denoted as 'Average Token2Statement', calculates the average value of the embedding of tokens within a statement. The specific process, illustrated in Fig.~\ref{fig:Token2Statement1}, is as follows: First, we construct a correspondence matrix $C \in \mathbb{R}^{L \times n}$ between statements and tokens. In this matrix, each row sequentially corresponds to a statement in the program, and each column corresponds to a specific token element in the token sequence. If the token element corresponding to a column is contained within the statement of a row, that position in the matrix is assigned the value 1; otherwise, it is 0. Next, we vertically concatenate all token vectors output by the TETransformer to form the Token matrix $T \in \mathbb{R}^{n \times d_h}$, where each row represents the vector representation of a token in the code sequence. Finally, by multiplying matrix $C$ with matrix $T$, we obtain the initial vector representation of each statement.
 
The process of initializing the aforementioned statement vectors essentially involves averaging the vector representations of all tokens contained in a statement. For instance, if a statement $c_i$ includes tokens from $w_o$ to $w_p$, its initial vector representation $\bm{s_i}$ can be computed using the following formula.

\begin{equation}
	\bm{s_i} =\frac{1}{|p-o|} \sum_{j \in [o \sim p]}^{} \bm{w_j}
    \label{eq:state_vector}
\end{equation}

\begin{figure}
	\centering{\includegraphics[width=0.5\textwidth]{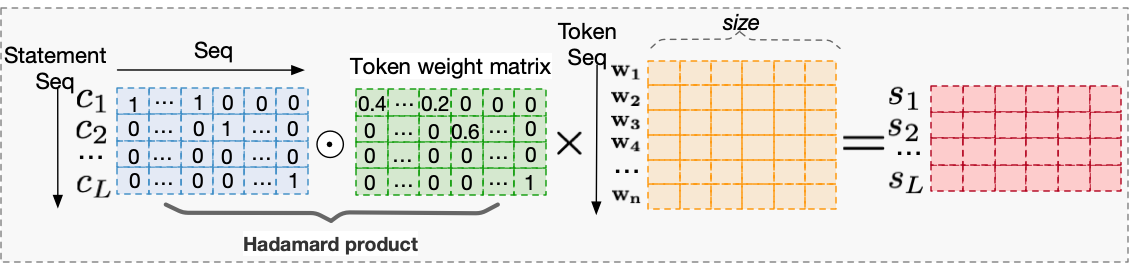}}
	\caption{Initializing statement vector representations using token vectors and statement-token correspondence, while accounting for the varied importance of tokens within statements}
	\label{fig:Token2Statement2}
\end{figure}

(2) The second strategy, referred to as 'Weighted Average Token2Statement', involves computing the weighted average value of the embeddings of all tokens within a statement. As indicated in formula~\ref{eq:state_vector}, the Average Token2Statement approach does not account for the varying importance of different tokens within a statement. For instance, API function calls in \textit{CallExpression} statements may carry more vulnerability information and thus hold greater importance~\cite{li2020vuldeelocator}. Recognizing this, we propose that when generating the vector representation of a statement from token vectors, the differing impacts of individual tokens should be considered. To achieve this, we introduce a learnable token weight matrix that adaptively fuses the semantic information of different tokens, thereby producing a more accurate initial vector representation for each statement. The process of Token2Statement embedding with the introduction of a weight matrix is illustrated in Fig.~\ref{fig:Token2Statement2}.

(3) The third strategy, known as 'Attention Average Token2Statement', differs from 'Weighted Average Token2Statement' in that it focuses on the contribution of each token to the representation of the entire code snippet, rather than solely relying on learnable token weights. In 'Attention Average Token2Statement', the weight assigned to each token is determined by its attention score relative to the [CLS] token, which is typically placed at the beginning of the token sequence to encapsulate the semantic information of the entire input code fragment. Specifically, this method first involves using a parameter matrix $\bf{Q}$ to map the [CLS] token vector to a query vector $q$, and a parameter matrix $\bf{K}$ to map each token vector to a key vector $k$. Then, the attention scores are computed as the product of $q$ and $k$. Finally, these attention scores are applied as weights to calculate the initial vector representation of each statement, which is achieved by performing a weighted summation of all token vectors contained within the statement.

\subsection{Statement Encode Transformer (SETransformer)} 
\label{dnet}
The SETransformer aims to capture the global semantic features between statements in a program, generating accurate vector representations for each line of code as well as for the entire code snippet. Similar to TETransformer, the SETransformer also employs a multi-headed self-attention mechanism to compute in parallel the degree of relevance between each statement and other dependent statements in the same code sequence. By doing so, it effectively establishes the complex semantic relationships between different statements, thereby achieving a comprehensive understanding and modeling of the program's global semantics.

\subsection{Algorithm for Processing Long Code Sequences}
\label{sec:algorithm}
In most pre-training settings for code language models, the maximum input sequence length is set to 512 tokens. However, this limitation prevents PCL models from scaling to process long and complex programs with vulnerabilities. To effectively tackle this challenge, we propose a long code encoding algorithm using CodeBERT-HLS, detailed in Algorithm 1. This algorithm is specifically designed to transform long code sequences into vector representations while preserving their semantic information.

\begin{algorithm}
	\caption{CodeBERT-HLS encodes long token sequence}
	\LinesNumbered 
	\small
    \SetKwInOut{Input}{Input}
	\SetKwInOut{Output}{Output}
\Input{Batch of samples, Maximum token sequence length \textit{m\_len} (e.g., 512, 1024, 2048)}
\Output{Vector representations of samples}
\For{sample in Batch}{
    \uIf{length(sample) $>$ m\_len}{
        sample = sample[:m\_len] //Truncate sample to max length\;
    }
    segment\_count =  $\left\lceil \frac{length(sample)}{512} \right\rceil$ \; 
    
    Split sample into segment\_count segments\;
    
    Initialize an empty list for token vectors: all\_token\_vectors = []\;
    \For{segment in segments}{
        token\_vectors = \textbf{TETransformer}(segment)\;
        Append token\_vectors to all\_token\_vectors\;
    }
    merged\_token\_vectors = Merge(all\_token\_vectors)\;
    state\_initial\_vectors = \textbf{Token2Statement}(merged\_token\_vectors)\;
    state\_vectors = \textbf{SETransformer}(state\_initial\_vectors)\;
    Output vector representations of each statement and the entire program\;
}
\end{algorithm}

Specifically, we assess whether the length of the input code sample exceeds the predefined maximum token sequence length, denoted as \textit{m\_len}. If it does, we truncate the sample (Lines 2-3). Then, we calculate the number of segments (i.e., segment\_count) required for splitting (Line 4). Subsequently, the sample is divided into segment\_count segments, and each segment is processed through the TETransformer to obtain the token vector representations (Lines 5-9). Next, we concatenate the token vectors from all segments and use Token2Statement to generate the initial vector representation for each line of code (Lines 10-11). Finally, these initial statement vectors are fed into the SETransformer to obtain the final statement vector representation (Lines 12). In this algorithm, the process of splitting each sample and handling each segment dynamically depends on the length of the sample and the parameter for maximum length. Therefore, this approach offers a more flexible way to handle code samples of varying lengths and effectively preserves the semantic information of the original samples.

Although Algorithm 1 processes individual samples for ease of understanding, in our implementation, we accelerate the generation of token vector representations for all samples in a batch by inputting their split code segments collectively into the TETransformer. 
Then, in the Merge function, segments belonging to the same sample are consolidated to form a matrix of initial token vectors. Finally, these token matrices for all samples in the batch are fed into Token2Statement and SETransformer to obtain vector representations of the statements.

\subsection{Pre-Training CodeBERT-HLS}
\label{sec:pre-training}
This section focuses on developing a pre-training task to devise more effective feature representations. This foundational work is essential for preparing the model to further specialize in accurately understanding and identifying code vulnerabilities through fine-tuning on labeled datasets. In this paper, we introduce the Masked Statement Prediction (MSP) model, a novel approach designed for effectively pre-training CodeBERT-HLS. Specifically, we select 15\% of the code lines from the input code fragment and apply the following transformations: in 80\% of these selected lines, we replace all tokens with an equal number of [MASK] tokens; in 10\% of the lines, we replace the tokens with random token sequences; and we leave the remaining 10\% of the lines unchanged. These numbers are chosen to align with the previous work, which commonly employ MLM for pretraining existing PCL models~\cite{feng2020codeBERT,guo2020graphcodeBERT}.
The objective of the MSP is to predict the original token sequences in these sampled code lines.

\begin{equation}
\mathcal{L}_{M S P}(\theta)=-\sum_{s \in S_m}\log P_\theta(\underbrace{x_i, \ldots x_j}_s \mid \mathbf{x}^{\text {mask }})
\end{equation}

\begin{figure}
	\centering{\includegraphics[width=0.5\textwidth]{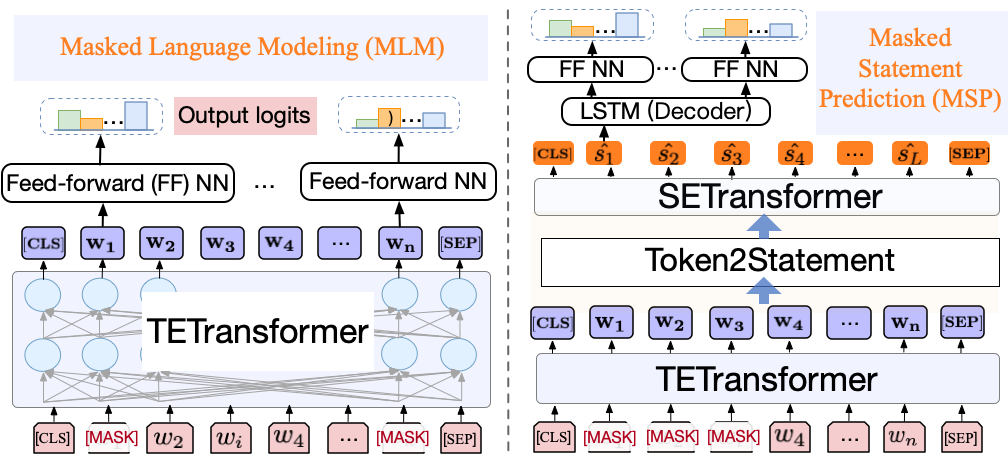}}
	\caption{Comparing the proposed MSP with the existing pre-training task MLM}
	\label{fig:MSP_MLM}
\end{figure}

Here, \(\theta\) denotes the model parameters, \(S_m\) is the set of selected masked statements, \(\{x_i, \ldots, x_j\}\) represents the token sequence within the masked statement \(s\), and \(\mathbf{x}^{\text{mask}}\) is the masked input code sequence. Given that the TETransformer adopts the same architecture as CodeBERT, we can initially pre-train the TETransformer using MLM and then jointly pre-train both the TETransformer and SETransformer with MSP. 
The comparison between MSP and MLM is illustrated in Fig.~\ref{fig:MSP_MLM}. 
As shown, MSP employs an LSTM to predict the token sequence of a masked statement based on its feature vector, whereas MLM uses an FFNN to predict the masked tokens based on the token feature vectors generated by the TETransformer. 

\subsection{Design Rationale of Our Method}
The design rationale behind our proposed CodeBERT-HLS is to capture both token-level and statement-level semantic information in source code, and make it suitable for multi-granular vulnerability detection. 
To achieve this, we introduce two key components: TETransformer and SETransformer.
The TETransformer focuses on learning token-level semantics, treating each token as the smallest semantic unit, whereas the SETransformer targets capturing statement-level semantics, considering each statement as the fundamental unit.
To effectively transfer semantic information from SETransformer to TETransformer, we develop three Token2Statement strategies. These strategies ensure a seamless connection between the two components, forming a cohesive and comprehensive architecture.
A notable advantage of our architecture is that it produces an accurate representation of each statement, which provides the basis and opportunities to develop an automatic fine-grained vulnerability detection model.
Based on the novel architecture, we design Algorithm 1 to enable the proposed architecture to efficiently process long token sequences. In addition, we introduce a new pre-training task to help our architecture learn general program semantics, which gives it the ability to adapt quickly to vulnerability detection tasks and achieve better generalization in real-world scenarios.

\section{Evaluation Setting}
\label{evaluation}
\subsection{Research Questions}
To evaluate the effectiveness of our method\footnote{
All resources, including code, data, and models, are publicly available at \url{https://github.com/YuanJiangGit/StagedVulBERT.git}}, we pose the following research questions (RQs). 
Given the significant contributions of CodeBERT-HLS and its pretraining task, we first assess their effectiveness in RQ1 and RQ2, respectively. In RQ3, we examine different components within CodeBERT-HLS. Next, in RQ4 and RQ5, we compare our approach's performance against state-of-the-art DL, transformer-based, and LLM-based methods in coarse-grained vulnerability detection. Finally, we demonstrate StagedVulBERT's significant performance enhancement in fine-grained vulnerability detection.
The specific RQs are as follows:

RQ1: To what extent does the proposed CodeBERT-HLS demonstrate statistically significant superiority over CodeBERT in vulnerability detection?

RQ2: To what extent is the proposed pre-training task effective in improving model performance?

RQ3: To what extent does each component of the proposed CodeBERT-HLS contribute to its overall effectiveness?

RQ4: To what extent does the proposed StagedVulBERT outperform the state-of-the-art DL- and transformer-based methods in coarse-grained vulnerability detection?

RQ5: To what extent does the proposed StagedVulBERT outperform the current state-of-the-art LLM-based methods in coarse-grained vulnerability detection?

RQ6: To what extent does the proposed StagedVulBERT effectively enhance the performance of fine-grained vulnerability detection?

\subsection{Dataset}
\label{dataset}
\begin{figure}
	\centering{\includegraphics[width=0.34\textwidth]{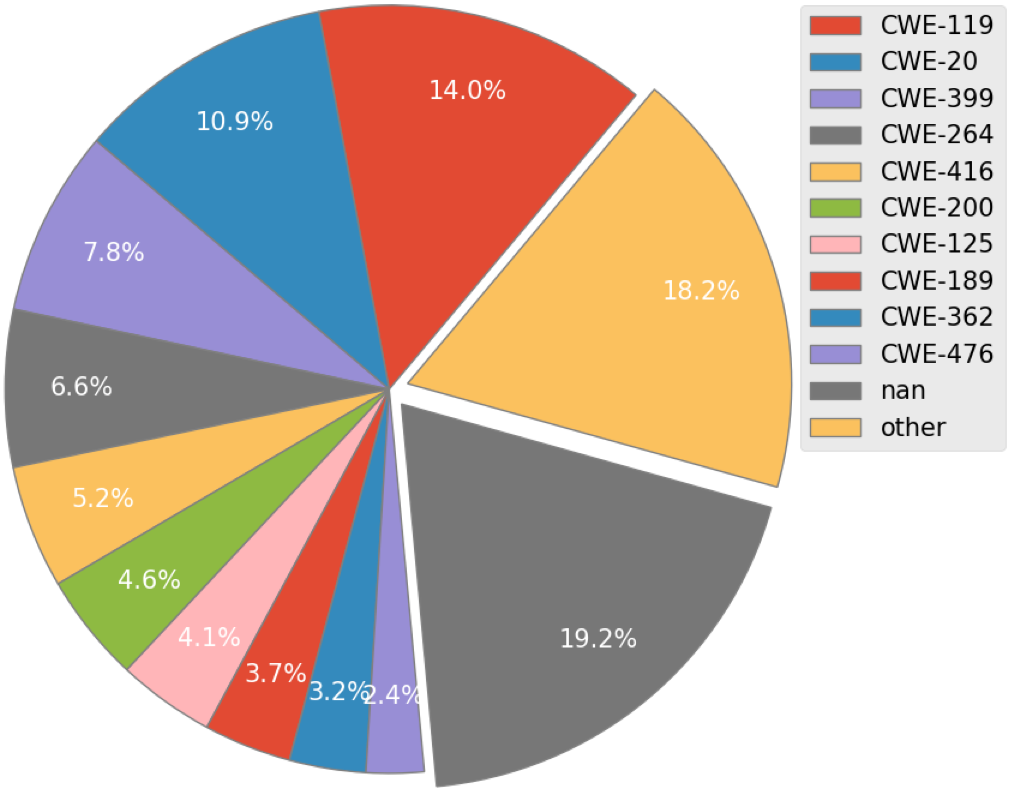}}
	\caption{The pie chart presents the distribution of vulnerability types of the programs in the dataset used in this paper.
	Each sector in the pie chart represents a different CWE whose size shows the proportion of programs with the specific type of vulnerabilities in the big\_vul dataset.}
	\label{fig:dataset-cwe-distribution}
\end{figure}

The dataset big\_vul we use in this paper is provided by Fan \textit{et al.}~\cite{fan2020ac}, which is collected from 348 open-source projects. This dataset is notable for its data instances from various real-world application domains, including web browsers like Chrome, operating systems such as Linux, mobile platforms like Android, and software applications such as FFmpeg.Concretely, big\_vul contains 188,636 samples, of which 10,900 are vulnerable functions and 177,736 are non-vulnerable functions. 
The vulnerable functions in the dataset are written in C/C++ and contain a total of 91 types of vulnerabilities, where CWE-119, CWE-20, CWE-399, CWE-264, CWE-416, CWE-200, CWE-125, CWE-189, CWE-362 and CWE-476 account for more than 60\% of the entire data. The vulnerability categorization is based on the Common Weakness Enumeration (CWE) system. 
The distribution of vulnerabilities across different CWE types is shown in Fig.~\ref{fig:dataset-cwe-distribution}.
To prepare the data for experiments, we follow the approach used in previous work~\cite{fu2022linevul} by randomly splitting the data into 80\% for training, 10\% for testing, and 10\% for evaluation. Table~\ref{cwe-num} shows the statistics of the programs in the dataset, and Table~\ref{pos-neg-num} details the number of vulnerable and non-vulnerable programs in training, evaluation, and testing datasets.

\begin{table}[htbp]
	\caption{The statistics of the dataset used in this paper, where \# iNum represents the number of programs in the dataset, \# MeaL, \# MinL, \# MaxL, \# MedL and \# StdL denote the mean, minimum, median, maximum and standard deviation of the number of lines of code in the programs, respectively.}
	\centering
	\resizebox{\columnwidth}{!}{
		\begin{tabular}{lrrrrrr}
			\toprule
			CWE ID & \# iNum & \# MeaL & \# MinL & \# MaxL & \# MedL & \# StdL  \\
			\midrule
            CWE-119 & 26,497 &461.1&5&115,726&178.0&1,275.3\\
            CWE-20 & 20,501 & 378.3&5&137,223&144.0&1,397.4\\
            CWE-399 & 14,806 & 309.2&5&50,427&132.0&1,124.3\\
            CWE-264 & 12,487 & 280.7&6&16,187&146.0&474.2\\
            CWE-416 & 9,780& 317.0&6&58,993&151.0&872.7\\
            CWE-200 & 8,636& 444.3&7&53,259&196.0&1,107.0\\
            CWE-125 & 7,717 & 630.6&5&44,158&236.0&1,552.8\\
            CWE-189 & 6,964 & 392.2&8&27,199&179.0&950.9\\
            CWE-362 & 6,073 & 344.8&5&30,443&170.0&735.7\\
            CWE-476 & 4,572 & 498.0&9&45,884&226.5&1,347.9\\
            \midrule
            ALL & 188,636 & 382.2&5&140,064&156.0&1202.5\\
			\bottomrule
		\end{tabular}
	}
	\label{cwe-num}
\end{table}

\begin{table}[htbp]
	\caption{The number of vulnerable and non-vulnerable programs in training, evaluating, and testing datasets}
	\centering
	\begin{tabular}{lrrrr}
		\toprule
		Dataset & \# Total & \# Vul & \# Non-Vul &\# Vul(\%)   \\
		\midrule
		Training Set & 150,908 & 8,736 & 142,172 & 5.79 \\
		Evaluation Set & 18,864 & 1,109& 17,755 & 5.88\\
		Testing Set & 18,864 & 1,055 &17,809 & 5.59\\
		\midrule
		Total & 188,636 & 10,900 & 177,736 & 5.78 \\
		\bottomrule
	\end{tabular}
	\label{pos-neg-num}
\end{table}

\subsection{Ground-truth Labels}
\label{labelling}
The big\_vul dataset from Fan \textit{et al.}~\cite{fan2020ac} is among the largest vulnerability datasets available, with each code function example labeled for both function-level and line-level ground truths. In this dataset, a code function within a project is labeled as vulnerable if it is modified in a vulnerability-fixing commit, which is recorded in a CVE entry. As in previous work~\cite{fu2022linevul}, these modified lines are considered as line-level labels for fine-grained vulnerability detection. Conversely, functions not modified in such commits are labeled as '0', indicating that they contain no vulnerabilities. The authors of big\_vul have invested considerable manual effort in the data collection process to ensure the high quality of this dataset~\cite{hin2022linevd}.

\subsection{Baseline Methods}
\label{baseline-methods}
\subsubsection{Methods for Coarse-grained Vulnerability Detection}

To demonstrate the superiority of the proposed method, we consider the following types of approaches as baselines:

(1) \textbf{Token-based methods} first transform code snippets or code slices into sequences of tokens which are then fed into CNN- or RNN-based models for vulnerability detection. Frequently cited comparative methods in this category include the one by Russell \textit{et al.}~\cite{Russell}, VulDeePecker~\cite{li2016vulpecker} and
SySeVR~\cite{li2018sysevr}.

(2) \textbf{GNN-based methods}, including notable approaches such as Devign~\cite{zhou2019devign}, Reveal~\cite{chakraborty2021deep}, IVDetect~\cite{li2021vulnerability}, and AMPLE~\cite{wen2023vulnerability}, construct various code graphs and then utilize graph neural networks (GNNs) to learn vulnerability patterns from these graphs. Empirical studies~\cite{chakraborty2021deep,wen2023vulnerability} have demonstrated that GNN-based methods outperform Token-based methods in vulnerability detection.

(3) \textbf{Transformer-based methods} employ pre-trained models to capture deep semantic features of code, making them particularly suitable for understanding complex vulnerabilities. Recent examples of such methods, LineVul~\cite{fu2022linevul}, SVulD~\cite{ni2023distinguishing} and VulBERTa~\cite{hanif2022vulBERTa}, have demonstrated promising results in identifying vulnerabilities within real-world datasets.

(4) \textbf{LLM-based methods} are trained with billions of parameters and training samples, which can also be explored in detecting vulnerabilities~\cite{fu2023chatgpt,yin2024multitask}. In comparison, the recently proposed models such as DeepSeek-Coder~\cite{guo2024deepseek}, CodeLlama~\cite{roziere2023code}, StarCoder~\cite{li2023starcoder}, WizardCoder~\cite{luo2023wizardcoder}, Mistral~\cite{jiang2023mistral}, Phi-2~\cite{javaheripi2023phi}, and ChatGPT~\cite{ouyang2022training} serve as baselines.

\subsubsection{Methods for Fine-grained Vulnerability Detection}
\label{sec:fine-baselines}
In this paper, we consider the following two types of fine-grained vulnerability detection methods as our baselines:

(1) \textbf{GNN-interpretation based methods} employ pre-trained GNN-based detection models to identify crucial features (e.g., nodes) for detection by assigning them importance scores, thereby achieving statement-level vulnerability prediction. In this paper, we select four GNN detectors (i.e., DeepWukong~\cite{cheng2021deepwukong}, Devign~\cite{zhou2019devign}, IVDetect~\cite{li2021vulnerability}, and Reveal~\cite{chakraborty2021deep}) and six GNN interpreters (e.g., GNN-LRP~\cite{schnake2021higher}, DeepLIFT~\cite{shrikumar2017learning}, GradCam~\cite{selvaraju2017grad}, GNNExplainer~\cite{ying2019gnnexplainer}, PGExplainer~\cite{luo2020parameterized}, and SubGraphX~\cite{yuan2021explainability}) collectively representing a wide spectrum of methods frequently used in this field. By pairing each detector with each interpreter, we consequently examine 24 distinct GNN-interpretation-based methods.

(2) \textbf{Attention-based methods}, such as LineVul~\cite{fu2022linevul}, calculate the total score for a statement by summing the attention scores of all its tokens. This score serves as a criterion to evaluate the semantic importance of the statement line. These methods then rank statement lines based on their attention scores, under the assumption that lines with higher ranks play a decisive role in the model's decision-making process.

\subsection{Parameter Configurations}
\label{sec:parameter}
In this paper, we pre-train CodeBERT-HLS using the widely-used CodeSearchNet dataset~\cite{husain2019codesearchnet}, which encompasses six different programming languages. Both the TETransformer and SETransformer in CodeBERT-HLS consist of six Transformer layers, each with a 768-dimensional hidden state and 12 self-attention heads. This configuration aims to facilitate a fair comparison with the 12-layer pre-trained CodeBERT model. Our pre-training is conducted on a computational platform equipped with five NVIDIA A100 GPUs. For pre-training StagedVulBERT, we adopt the following hyperparameter settings by conducting a random search in the hyperparameter space: a batch size of 400, a learning rate of 5e-5, a maximum input length of 512 tokens, and a maximum of 5 training epochs. Additionally, the AdamW optimizer is used to update the model parameters. 
Training 100K batches of data with MSP takes approximately 130 hours. Due to the high cost associated with searching for optimal hyperparameters on the entire CodeSearchNet dataset, we only validate these parameters on a small subset of the data to ensure they yield satisfactory performance compared to other parameter combinations.

After pre-training, we further fine-tune StagedVulBERT, integrating a classification network for vulnerability detection, using the big\_vul dataset. This fine-tuning is performed on a server equipped with an A6000 (48G) GPU. The hyperparameters for fine-tuning are set as follows: a batch size of 32, a range of maximum input lengths among \{512, 1024, 2048\} tokens, and a learning rate of 2e-5. Similar to the pre-training phase, the AdamW optimizer is employed to update the model parameters. Completing a full fine-tuning cycle involves 10 epochs and takes approximately 10 hours. Note that we adopt a smaller batch size compared to pre-training, making it easier for readers to reproduce the fine-tuned model presented in this paper.

\subsection{Evaluation Metrics}
\label{metrics}
To evaluate the performance of our proposed method, we use the following five metrics, which have been widely accepted by previous work~\cite{li2018sysevr,hu2023interpreters}.

\begin{itemize}
	\item Precision (precision) is the fraction of predicted vulnerabilities that are correctly predicted~\cite{davis2006relationship,powers2011evaluation}; $ precision= \frac{TP}{TP+FP} $.
	\item Recall (recall) is the fraction of true positive vulnerabilities in the actual vulnerabilities~\cite{davis2006relationship,powers2011evaluation}; $ recall= \frac{TP}{TP+FN} $.
	\item F1 score is used to balance precision and recall. It is an aggregate way to describe a model~\cite{powers2011evaluation}; $ \textit{F1 score}=2 \times \frac{precision \times (1-FNR).}{precision+(1-FNR)} $
	\item Accuracy (accuracy) is the percentage of correct predictions: $ accuracy=\frac{TP+TN}{TP+FP+TN+FN} $·
	\item Top-\textit{k}\% Accuracy (Top-\textit{k}\% Acc) is the locating accuracy of a vulnerability detector $e$. It measures whether the top \textit{k}\% of statements predicted as vulnerable by the detector are correct by checking if they overlap with the fine-grained ground truth. The formula is as follows, where $r_i$ denotes the fine-grained label of sample $x_i \in D$ and $p_i$ is the prediction of the top \textit{k}\% most vulnerable statements.
     \begin{equation}
    \begin{gathered}
    A c c(e, D, k)=\frac{1}{|D|} \sum_{x_i \in D} A c c\left(e, x_i, k\right) \\
    \operatorname{Acc}\left(e, x_i, k\right)=\left\{\begin{array}{l}
    1, r_i \cap p_i \neq \emptyset \\
    0, r_i \cap p_i=\emptyset
    \end{array}\right.
    \end{gathered}
    \end{equation}
\end{itemize}

\section{Evaluation Results}
\label{results}

\subsection{Experiments for Answering RQ1}
\label{rq1}
CodeBERT, as employed in the recently proposed LineVul method by Fu et al.~\cite{fu2022linevul} and the VulBERTa method by Hanif et al.~\cite{hanif2022vulBERTa}, has been successfully applied to the task of vulnerability detection, achieving satisfactory performance. Given that both our proposed model CodeBERT-HLS and CodeBERT are based on a transformer-encoder architecture and adhere to a similar learning protocol of extensive pre-training on large datasets via self-supervised learning followed by fine-tuning for vulnerability detection, this section aims to determine whether our model exhibits a statistically significant advantage over CodeBERT. 

To address these questions, we randomly divide the big\_vul data into ten disjoint, equal-sized sets, named $fold_{1}$, $fold_{2}$, ..., $fold_{10}$. We then conduct cross-validation, training our method on $fold_{i}$, evaluating it on $fold_{i+1}$, and testing it on $fold_{i+2}$. Since our method accommodates input sequences of varying token lengths, we also compare different variants of our approach, specifically StagedVulBERT (512), StagedVulBERT (1024), and StagedVulBERT (2048). These variants employ the same network architecture, CodeBert-HLS with the 'Average Token2Statement', but differ only in the length of input samples processed, with specific token counts of 512, 1024, and 2048, respectively.
For each method, we obtain eight values of accuracy, recall, precision and F1 score corresponding to the eight testing sets. 

\begin{figure*}
	\centering{\includegraphics[width=1.0\textwidth]{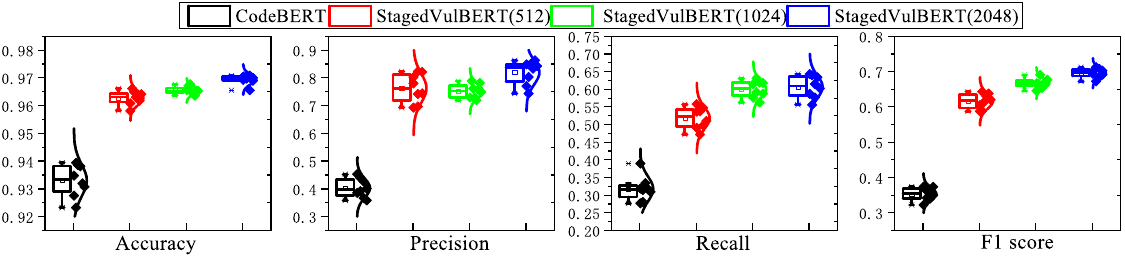}}
	\caption{Performance comparison of StagedVulBERT (with input lengths of 512, 1024, and 2048) and CodeBERT in coarse-grained vulnerability detection without pre-training using boxplots. Each boxplot represents the performance distribution of each method on split datasets. StagedVulBERT variants differ in their maximum input length: 512, 1024, and 2048. For example, StagedVulBERT (512) has a maximum input length of 512.}
	\label{fig:no_pretrain_compare}
\end{figure*}

Taking the overall effectiveness, as measured by the F1 score, as an example: without pre-training, our methods StagedVulBERT (512), StagedVulBERT (1024), and StagedVulBERT (2048) show an improvement over CodeBERT by 74.40\%, 88.87\%, and 96.97\% respectively in terms of average performance, as illustrated in Fig.~\ref{fig:no_pretrain_compare}. This suggests that our proposed model architecture is more suited for the vulnerability detection task compared to the CodeBERT model.

To determine whether our method demonstrates a statistically significant advantage, we perform a two-tailed Wilcoxon Signed-Rank Test~\cite{wilcoxon1992individual} to analyze the differences between pairs of results from each sub-figure in Figs.~\ref{fig:no_pretrain_compare}. Recall that a significance level of 0.05 or less (indicating 95\% confidence) denotes statistical significance. We observe that all the \textit{T} statistic for Wilcoxon's test ($ T_{wilcox} $) are smaller than the critical value as provided by~\cite{japkowicz2011evaluating}, suggesting that our model architecture has a statistically significant advantage.

\textbf{Conclusion}: Without pre-training, StagedVulBERT with CodeBERT-HLS outperforms LineVul with CodeBERT, as CodeBERT-HLS has a more suitable network architecture for vulnerability detection tasks.

\subsection{Experiments for Answering RQ2}
\begin{figure*}
	\centering{\includegraphics[width=1.0\textwidth]{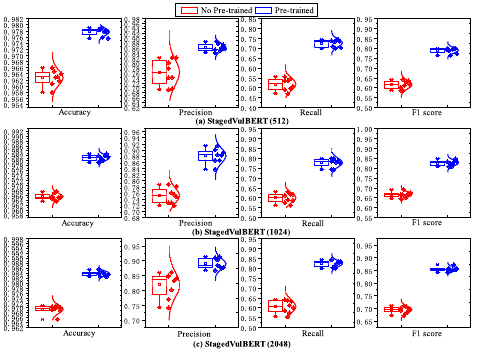}}
	\caption{Performance comparison of Pre-trained StagedVulBERT versus No Pre-trained StagedVulBERT in coarse-grained vulnerability detection using boxplots. The first row of subfigures presents experiments conducted with a 512 input length limit, while the second and third rows display the results for the 1024 and 2048 input lengths, respectively. Each boxplot represents the performance distribution of each method on the split datasets.
 }
	\label{fig:pretrain_or_not}
\end{figure*}

RQ1 has demonstrated the effectiveness of the proposed CodeBERT-HLS; this section investigates the role of the pre-training task introduced in this paper in enhancing CodeBERT-HLS's capability for vulnerability detection. To achieve this, we use the steps in Section~\ref{sec:pre-training} and the hyper-parameters detailed in Section~\ref{sec:parameter} to pre-train CodeBERT-HLS. We then compare the performance differences of StagedVulBERT with the pre-trained or non-pre-trained CodeBERT-HLS in coarse-grained vulnerability detection.
Finally, we perform cross-validation with ten equally sized subsets from $fold_{1}$ to $fold_{10}$, which is inconsistent with the experimental settings in RQ1.
We also investigate the effect of varying input sequence lengths (512, 1024, 2048) on the performance of both pre-trained and non-pre-trained CodeBERT-HLS models. The experimental results are provided in Fig.~\ref{fig:pretrain_or_not}.

As shown in Fig.~\ref{fig:pretrain_or_not}, when initialized with pre-trained weights, StagedVulBERT (512), StagedVulBERT (1024), and StagedVulBERT (2048) achieve average F1 scores of 78.94\%, 82.63\%, and 85.50\% respectively across eight testing sets. These scores represent improvements of 28.19\%, 23.91\%, and 22.94\% over the variants without pre-training. This indicates that the proposed pre-training task MSP effectively assists the model in learning more generalized feature representations of programs, thereby enhancing performance in downstream vulnerability detection.

Furthermore, to determine whether the performance improvements are statistically significant, we conduct the standard Wilcoxon Signed-Rank Test at a 0.05 significance level for pairs of results from each sub-figure in Fig.~\ref{fig:pretrain_or_not}. We find that all T-statistics for Wilcoxon's test ($T_{\text{Wilcox}}$) are below the critical value, as per \cite{japkowicz2011evaluating}, indicating that our pre-training task is effective and the results are statistically significant.
Given that StagedVulBERT (2048) with pre-training shows the best performance, it has been selected as our primary method for subsequent experiments.

\textbf{Conclusion}: With pre-training, StagedVulBERT with CodeBERT-HLS exhibits significantly enhanced effectiveness compared to its version without pre-training. This emphasizes the importance of pre-optimization through the proposed MSP task to achieve the best performance.

\subsection{Experiments for Answering RQ3}
The effectiveness of the entire architecture of CodeBERT-HLS has been demonstrated previously in RQ1. In this section, we will investigate the effectiveness of each component (i.e., TETransformer, Token2Statement Embedding, and SETransformer) separately. 
The TETransformer, sharing the same architecture with CodeBERT, can be directly utilized for vulnerability detection without modification. However, the SETransformer, which takes the initial representation of each statement as input, requires adaptation for vulnerability detection. To achieve this, the average embedding of all tokens within a statement is employed as input. In addition, relying solely on Token2Statement is insufficient for direct vulnerability detection, as it is situated in an intermediate layer and cannot directly produce predictive output. Thus, we analyze the performance of different design choices for Token2Statement embedding with the same SETransformer and TETransformer. The detailed experimental results of our method's variants with pre-training are presented in Table~\ref{token2statement_impact}.

\begin{table}[ht]
\centering
\caption{Performance comparison of our method's variants: TETransformer (TE), SETransformer (SE), and their combination with three Token2Statement strategies—Average (Ave), Weighted Average (Wei), and Attention Average (Alt) Token2Statement.}
\begin{tabular}{lcccc}
\toprule
\textbf{Approach} & \textbf{Accuracy} & \textbf{Precision} & \textbf{Recall} & \textbf{F1 score} \\
\midrule
\multicolumn{5}{c}{Only use TETransformer (TE) or SETransformer (SE)} \\
\midrule
TE& 97.84 &88.03 & 71.09 & 78.66 \\
SE & \textbf{97.92} & 86.94 & \textbf{73.84} & \textbf{79.86} \\
\midrule

\multicolumn{5}{c}{TE and SE with different Token2Statement} \\
\midrule
TE+SE+Ave & \textbf{99.15} & \textbf{94.26} & 90.33 & \textbf{92.26} \\
TE+SE+Wei & 99.12 & 93.88 & 90.14 & 91.97 \\
TE+SE+Att & 99.10 & 92.59 & \textbf{91.18} & 91.88 \\
\bottomrule
\end{tabular}
\label{token2statement_impact}
\end{table}

As can be seen from Table~\ref{token2statement_impact}, when employing solely the TETransformer or SETransformer, a strong drop in detection performance is observed due to reduced program comprehension ability.
Additionally, combining TETransformer and SETransformer with ``Average Token2Statement" achieves a better performance than the combination with ``Weighted Average Token2Statement".
This indicates that determining the semantic importance of a token based on its absolute position during Token2Statement stage does not enhance the model's code representation capability. This is because positional embeddings in the TETransformer sufficiently capture token locations, which are vital for semantic understanding in subsequent layers. Adding positional weighting in the Token2Statement embedding does not consistently boost performance; thus, the ``Weighted Average Token2Statement" method, duplicating absolute position capture, offers no benefits.

Among all the variants, the one employing the ``Attention Average Token2Statement" strategy shows the poorest performance.
One possible explanation is that the attention scores might not accurately reflect the true significance of tokens for the given task. This finding is consistent with similar research, such as LineVul, which also tried to use token attention scores for identifying potentially vulnerable statements but faced limitations. As demonstrated in Section 2.1, relying solely on token attention scores fails to precisely pinpoint vulnerabilities at a fine-grained level. Hence, it appears that at the present scale of training data, the attention scores output by the PCL model for tokens may not accurately represent the semantic importance of these tokens, and could even lack a direct correlation with their semantics. This observation helps to understand why the ``Attention Average Token2Statement" variant falls short in performance when compared to others.

\textbf{Conclusion}: Our investigation into StagedVulBERT's components demonstrates their effectiveness in detecting vulnerabilities. Combining these components achieves optimal performance, particularly when employing the ``Average Token2Statement" in the intermediate layer.

\subsection{Experiments for Answering RQ4}
In addition to demonstrating the effectiveness of the model architecture and pre-training task discussed in RQ1 and RQ2, respectively, we also aim to assess whether our method outperforms current state-of-the-art approaches in coarse-grained vulnerability detection.
To answer this question, we compare our approach with ten baselines including three RNN-based methods, four GNN-based methods and three transformer-based methods. Our StagedVulBERT approach, with a 2048 input token limit, is pre-trained on CodeSearchNet and fine-tuned on the training set of big\_vul.
Table~\ref{rq1-result} shows the experimental results on the big\_vul dataset.

\begin{table}[htbp]
    \caption{Comparisons between our method and others in coarse-grained vulnerability detection when evaluated on the big\_vul dataset}
    \centering
    \resizebox{\columnwidth}{!}{
    \begin{tabular}{@{}llllll@{}}
        \toprule
        \textbf{Type} & \textbf{Baseline} & \textbf{Accuracy} & \textbf{Precision} & \textbf{Recall} & \textbf{F1 score} \\ 
        \midrule
        \multirow{3}{*}{\makecell[l]{RNN-based\\methods}} & VulDeePecker & 87.19 & 38.44 & 12.75 & 19.15 \\
        & Russell \textit{et al.} & 86.85 & 14.86 & 26.97 & 19.17 \\
        & SySeVR & 90.10 & 30.91 & 14.08 & 19.34 \\ 
        \midrule
        \multirow{4}{*}{\makecell[l]{GNN-based\\methods}} & Devign & 92.78 & 30.61 & 15.96 & 20.98 \\
        & Reveal & 87.14 & 17.22 & 34.04 & 22.87 \\
        & IVDetect & - & 23 & 72 & 35 \\
        & AMPLE & 93.14 & 29.98 & 34.58 & 32.11 \\ 
        \midrule
        \multirow{4}{*}{\makecell[l]{Transformer-based\\methods}} & VulBERTa & 93.99 & 48.44 & 33.54 & 39.64 \\
        & SVulD & 97.1 & 72.4 & 79.3 & 75.7 \\
        & LineVul & 98.53 & 94.2 & 78.58 & 85.68 \\
        & StagedVulBERT & \textbf{99.15} & \textbf{94.26} & \textbf{90.33} & \textbf{92.26} \\
        \bottomrule
    \end{tabular}
    }
    \label{rq1-result}
\end{table}

Table~\ref{rq1-result} shows that Transformer-based methods perform better than other DL-based models. This is consistent with the observations by previous work which confirms the powerful capability of Transformer-based methods for automatic vulnerability feature extraction and representation~\cite{fu2022linevul}. Among existing methods, LineVul exhibits the best performance with an F1 score of 85.68\%, though this falls short of the 91.0\% reported in the original paper. This discrepancy is due to the original implementation of LineVul removing all code indentation from the vulnerable programs in the big\_vul dataset during its data preprocessing phase. This leads to a format difference between vulnerable and non-vulnerable code, thereby reducing the difficulty of identifying vulnerable code and resulting in a higher F1 score.
To more accurately reflect real-world scenarios, we address this issue and re-conduct experiments on the original dataset using the publicly available implementation of LineVul without such preprocessing.

Compared to the state-of-the-art Transformer-based methods, StagedVulBERT demonstrates notable performance improvements. It achieves an 11.75\% increase in recall and a substantial 6.58\% improvement in F1 score over LineVul. While accuracy sees a slight increase of 0.62\%, the overall performance gain is evident, particularly in the critical metrics of recall and F1 score, which are vital for a comprehensive evaluation of vulnerability detection models.
The higher detection performance could be due to the fact that the proposed CodeBERT-HLS used in StagedVulBERT is more capable of learning semantic and syntactic of code, particularly for larger programs, compared to other methods (e.g., LineVul and SVulD) that are limited to processing input sequences of no more than 512 tokens.

\begin{figure}
	\centering{\includegraphics[width=0.5\textwidth]{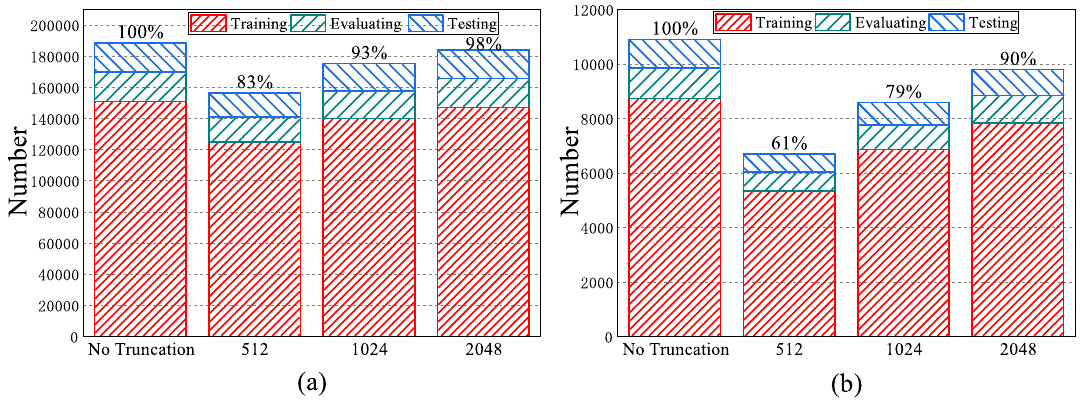}}
	\caption{Number of untruncated samples (Vertical axis) in the big\_vul dataset at different truncation lengths (Horizontal axis). The left graph (a) shows statistics for the entire dataset, while the right graph (b) displays results specifically for the vulnerable data in the big\_vul.}
	\label{fig:truncat-distribution}
\end{figure}

To further investigate whether our method can effectively process a greater number of real-world programs compared to methods like LineVul, we conduct a statistical analysis examining the number of truncated samples under various input length constraints.
As illustrated in Fig.~\ref{fig:truncat-distribution} (left), with a 512-token limit, 17\% of samples in the big\_vul dataset are truncated, leading to potential information loss. However, increasing the token limit to 1024 and 2048 reduces truncation to 7\% and 2\%, respectively, thus preserving more of the original code information.
Moreover, when focusing specifically on the vulnerable functions in the big\_vul dataset (Fig.~\ref{fig:truncat-distribution} (right)), only 61\% of vulnerable programs are fully processed by the model when truncated at 512 tokens. This percentage increases to 79\% and 90\% when the input length is extended to 1024 and 2048 tokens, respectively. Consequently, with its ability to handle longer sequences, CodeBERT-HLS learns more comprehensive code features, particularly those from complex vulnerability code, thereby boosting its vulnerability detection capabilities.

\textbf{Conclusion}: Our method achieves a greater improvement than the state-of-the-art DL-based model, especially in the F1 score (from 85.68\% to 92.26\%). This could be due to the fact that our method can correctly learn code vector representations from the vulnerability dataset.

\subsection{Experiments for Answering RQ5}

LLMs have shown outstanding performance in various code-related tasks, such as code generation and code search. This section aims to explore whether LLMs also exhibit similar excellence in the vulnerability detection task. We also wonder whether our proposed method provides superior vulnerability prediction compared to those LLM-based methods. This exploration would help determine the necessity for further research into specific transformer-based methods and technologies.

To evaluate the performance differences between StagedVulBERT and LLMs in coarse-grained vulnerability detection, we conduct experiments on the big\_vul dataset. The experiments involve comparisons with six recently proposed open-source LLMs: DeepSeek-Coder 6.7B, CodeLlama 7B, StarCoder 7B, WizardCoder 7B, Mistral 7B, and Phi-2 2.7B, as well as two close-source LLMs from OpenAI: Chatgpt-3.5 and Chatgpt-4. To make these open-source LLMs align with the vulnerability detection task, we employ the "AutoModelForSequenceClassification" from the Transformers library\footnote{https://pypi.org/project/transformers/, a Python library} to implement discriminative vulnerability detection. This model class seamlessly appends a classification layer on top of each pre-trained LLM, adapting it for sequence classification tasks. We fine-tune this model using the ``big\_vul" dataset, enabling LLMs to accurately classify code inputs as ``vulnerable" or ``not vulnerable". 
The above experimental settings are consistent with recent state-of-the-art empirical research~\cite{yin2024multitask}. 
For ChatGPT models, we use a few-shot setting to construct prompts designed for vulnerability detection, as detailed in related work~\cite{fu2023chatgpt}.
The results of these experiments are presented in Table~\ref{tab:compare2llm}. From these, we can draw the following conclusions:

\begin{table}[htbp]
    \caption{Performance comparison of LLM-based methods and StagedVulBERT in coarse-grained vulnerability detection when evaluated on the big\_vul dataset}
    \centering
    \resizebox{\columnwidth}{!}{
    \begin{tabular}{@{}llllll@{}}
        \toprule
        \textbf{Type} & \textbf{Baseline} & \textbf{Accuracy} & \textbf{Precision} & \textbf{Recall} & \textbf{F1 score} \\ 
        \midrule
        \multirow{2}{*}{\makecell[l]{LLM-based\\(Business)}} & Chatgpt-3.5 & 0.907 & 0.115 & 0.912 & 0.103 \\
        &Chatgpt-4 & 0.911 & 0.251 & 0.274 & 0.292 \\
        \midrule
        \multirow{6}{*}{\makecell[l]{LLM-based\\(Open-source)}} & DeepSeek-Coder & 0.964 & 0.633 & 0.886 & 0.739 \\
        &CodeLlama & 0.958 & 0.598 & 0.850 & 0.702 \\
        &StarCoder & 0.955 & 0.579 & 0.784 & 0.666 \\
        &WizardCoder & 0.955 & 0.574 & 0.882 & 0.695 \\
        &Mistral & 0.919 & 0.405 & 0.873 & 0.553 \\
        &Phi-2 & 0.955 & 0.569 & 0.896 & 0.696 \\
        \midrule
        \multirow{1}{*}{\makecell[l]{Our methods}} & StagedVulBERT & \textbf{0.992} & \textbf{0.943} & \textbf{0.903} & \textbf{0.923} \\
        \bottomrule
    \end{tabular}}
    \label{tab:compare2llm}
\end{table}

Our proposed StagedVulBERT significantly outperforms LLM-based methods in terms of all evaluation metrics, with the latter showing relatively weaker performance. 
For instance, StagedVulBERT achieves an F1 score of 92.3\%, while the better-performing baseline (i.e., DeepSeek-Coder) only reaches 73.9\%.
In addition, LLMs fine-tuned for vulnerability detection perform better than those LLMs (e.g., Chatgpt-4) in the few-shot setting, as observed in~\cite{yin2024multitask}.
These results suggest that despite LLM's excellence in natural language processing, it still falls behind models specifically designed for vulnerability detection, such as StagedVulBERT. This also demonstrates the significance of continuing research into PCL models for the vulnerability detection task.

\textbf{Conclusion}: LLM can be used in detecting software vulnerabilities, but fine-tuned open-source LLMs and few-shot commercial LLMs perform weaker than our method.

\subsection{Experiments for Answering RQ6}
To demonstrate the superiority of our proposed method in fine-grained vulnerability detection, we compare it with existing fine-grained vulnerability detection baselines, as outlined in Section~\ref{sec:fine-baselines}, using the big\_vul dataset annotated with fine-grained labels. Our evaluation metric involves calculating the accuracy of the model's predictions for the top \textit{k}\% of code lines, with \textit{k} ranging from 2 to 20. The experimental results are depicted in Fig.~\ref{fine_grained_result}. From these results, we can draw the following conclusions:

\begin{figure*}[htbp]
	\centerline{
		\includegraphics[width=1.0\textwidth]{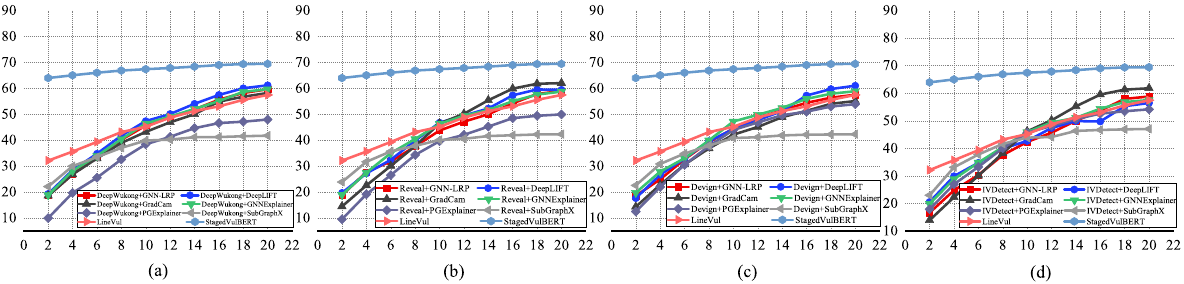}}
	\caption{Comparisons between our method and others in fine-grained vulnerability detection when evaluated on the big\_vul dataset. The horizontal axis represents the \textit{k} values ranging from 2 to 20, while the vertical axis shows the accuracy of the Top-\textit{k}\% predicted statements. Due to a total of 24 GNN-interpretation based methods in the baselines, which is too numerous to display in a single figure, we have evenly distributed them across four sub-figures: (a), (b), (c), and (d).}
	\label{fine_grained_result}
\end{figure*}

StagedVulBERT, proposed in this paper, significantly outperforms other methods in fine-grained vulnerability detection. When compared with GNN-interpretation-based methods and the attention-based method, StagedVulBERT consistently shows superior accuracy across various settings of \textit{k}. For instance, at \textit{k}=5, the best-performing baseline method, LineVul, achieves an accuracy of 37.5\%, whereas StagedVulBERT reaches an impressive 65.69\% under the same conditions, marking a performance increase of 75.17\%. This substantial improvement is largely due to the fact that previous studies focus on training models for coarse-grained detection using only coarse-grained labels via calculating maximum likelihood, and then relying on interpreters or attention scores to assess the contribution of each code line to the model's output. In contrast, StagedVulBERT, after initially training a coarse-grained vulnerability detection model, employs supervised learning to further train a model specifically designed for vulnerable statement classification. This allows it to directly determine the relevance of a statement to vulnerabilities based on its feature representation. By effectively leveraging both coarse- and fine-grained labels during training, StagedVulBERT significantly enhances its ability to precisely locate specific lines of vulnerable code.

Another factor contributing to StagedVulBERT's high performance is the proposed CodeBERT-HLS's ability to generate accurate statement feature representations. By employing the self-attention mechanism of the SETransformer, it captures the interdependencies between statements. This means that the generated statement features contain not only the semantics of the statement itself but also an integrated context of related information. Moreover, as the SETransformer can process up to 512 lines of statements, it captures a longer range of dependencies than other token-based models. Therefore, even though our statement classifier utilizes a simple feedforward neural network, it still achieves excellent experimental results. This approach, which considers both the local semantic information of statements and the overall structural dependencies in the code, provides stronger support for accurately locating and understanding vulnerabilities.

\textbf{Conclusion}: Compared to the baselines, our proposed approach demonstrates superior effectiveness in locating vulnerable lines of code, which achieves a higher locating precision compared to the state-of-the-art fine-grained vulnerability detection methods.

\section{Discussion}
\label{discussion}

\subsection{How Accurate is Our StagedVulBERT in Predicting the Most Frequent and Dangerous CWEs?}

To validate the practicality of StagedVulBERT in real-world scenarios, we conduct experiments on 14 types of CWE code samples, which are the most frequent and dangerous in the big\_vul dataset. Among these, CWE-119, CWE-20, CWE-399, CWE-264, CWE-416, CWE-200, CWE-125, CWE-189, CWE-362 and CWE-476 are the most commonly occurring CWE types in the big\_vul dataset, while CWE-787, CWE-125, CWE-20, CWE-416, CWE-22, CWE-190, CWE-476, CWE-119 and CWE-77 are recognized among the Top-25 most dangerous CWEs~\cite{CWE2023}.
For each of these 14 CWE types, we evaluate StagedVulBERT’s performance in both coarse- and fine-grained vulnerability detection. 

\begin{figure*}[htbp]
	\centerline{
		\includegraphics[width=1.0\textwidth]{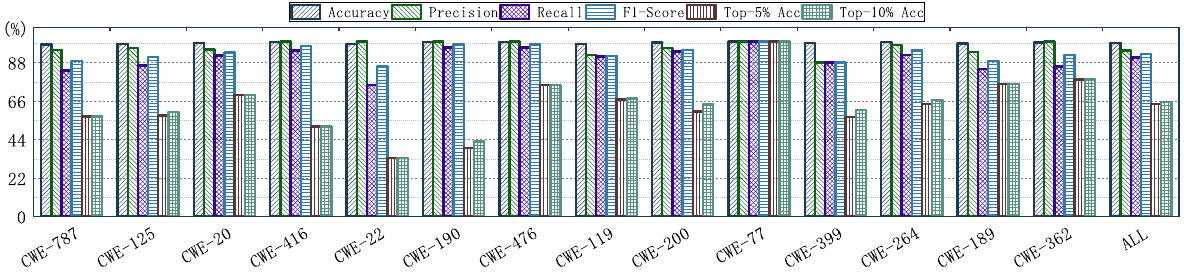}}
	\caption{The performance of our proposed method StagedVulBERT to detect vulnerabilities of real-world programs with the top-14 CWE vulnerability types.}
	\label{fig:CWE-Perform-Dist}
\end{figure*}

The results, presented in Fig.~\ref{fig:CWE-Perform-Dist}, show that in coarse-grained detection, StagedVulBERT achieves average accuracy, precision, and F1 score of 99.19\%, 94.95\%, and 92.81\%, respectively. Notably, StagedVulBERT shows weaker performance in detecting CWE-22 (path traversal) vulnerabilities, with an F1 score of only 85.71\%, but excelled at detecting CWE-77 (improper neutralization) vulnerabilities, achieving an F1 score of 100\%. In fine-grained detection, StagedVulBERT also demonstrates superior performance, with Top-5\% and Top-10\% accuracy rates of 64.33\% and 65.42\%, respectively. This further confirms its significant advantages in precisely identifying and locating specific vulnerabilities in code, particularly in scenarios involving a range of different vulnerability types.

\begin{figure}[htbp]
	\centerline{
		\includegraphics[width=0.5\textwidth]{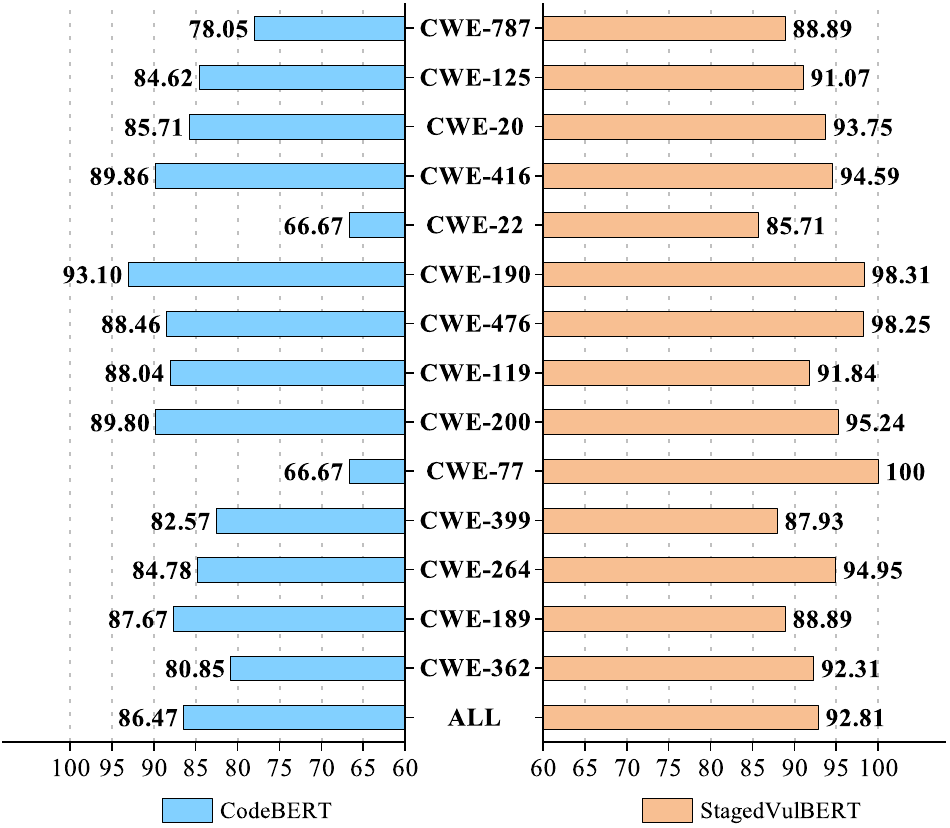}}
	\caption{The accuracy of our method and the state-of-the-art method for the Top-14 common and high-risk CWEs}
	\label{fig:CWE-Perform-compare}
\end{figure}

Furthermore, we present a comparison of the F1 score performance between the proposed StagedVulBERT and the state-of-the-art method, LineVul with CodeBERT, on the most frequent and dangerous vulnerability samples. As illustrated in Fig.~\ref{fig:CWE-Perform-compare}, StagedVulBERT achieves performance improvements in all these CWEs. For instance, the improvements in CWE-22 (path traversal) and CWE-77 (improper neutralization) vulnerabilities are 28.56\% and 50.0\%, respectively. Overall, StagedVulBERT's average F1 score on these 14 CWEs of samples shows a 7.33\% increase compared to CodeBERT.

\subsection{Investigate Whether the Proposed StagedVulBERT Can Generalize to Other Datasets}
The big\_vul dataset, widely used in recent studies due to its large size and real-world project origins, accurately reflects sample imbalance in actual scenarios, thus providing a more authentic experimental environment for vulnerability detection. Even with this, we would like to explore whether our method can generalize well to other datasets. To achieve this, we select a new dataset publicly published by~\cite{li2020vuldeelocator}, which consists of 14,511 programs, each labeled with detailed vulnerability information down to the line level. The experimental setting is the same as detailed in Section~\ref{sec:parameter} and the experimental results are shown in Table~\ref{generalize}.

\begin{table}[htbp]
    \caption{Comparisons between our method and LineVul when evaluated on the dataset used in \cite{li2020vuldeelocator}}
    \centering
    \resizebox{\columnwidth}{!}{
    \begin{tabular}{@{}ccccccc@{}}
        \toprule
        \multirow{2}{*}{\textbf{Model}} & \multicolumn{4}{c}{Coarse-grained} & \multicolumn{2}{c@{}}{Fine-grained} \\
        \cmidrule(lr){2-5} \cmidrule(l){6-7}
        & Accuracy & Precision & Recall & F1 score & Top-5\% Acc & Top-10\% Acc \\
        \midrule
        LineVul & 92.29 & \textbf{84.78} & 84.86 & 84.82 & 21.00 & 25.00 \\
        StagedVulBERT & \textbf{92.73} & 83.51 & \textbf{88.93} & \textbf{86.14} & \textbf{94.57} & \textbf{95.54} \\
        \bottomrule
    \end{tabular}
    }
    \label{generalize}
\end{table}

As can be seen in Table~\ref{generalize}, on the dataset published by~\cite{li2020vuldeelocator}, our proposed StagedVulBERT outperforms LineVul in both coarse- and fine-grained vulnerability detection tasks, with the latter particularly underperforming in fine-grained detection.
For instance, in fine-grained detection, StagedVulBERT's Top-10\% Accuracy is 95.54\%, significantly higher than the 25.00\% of the LineVul, indicating a substantial gap. 
These results suggest that the statement vectors generated by our proposed CodeBERT-HLS effectively capture the semantics of statements and are better suited as input for fine-grained vulnerability detection models compared to the token vectors produced by the baseline method.

\subsection{Investigate the Cases where StagedVulBERT Model Failed to Detect Vulnerabilities}
\begin{figure}[htbp]
	\centerline{
		\includegraphics[width=0.5\textwidth]{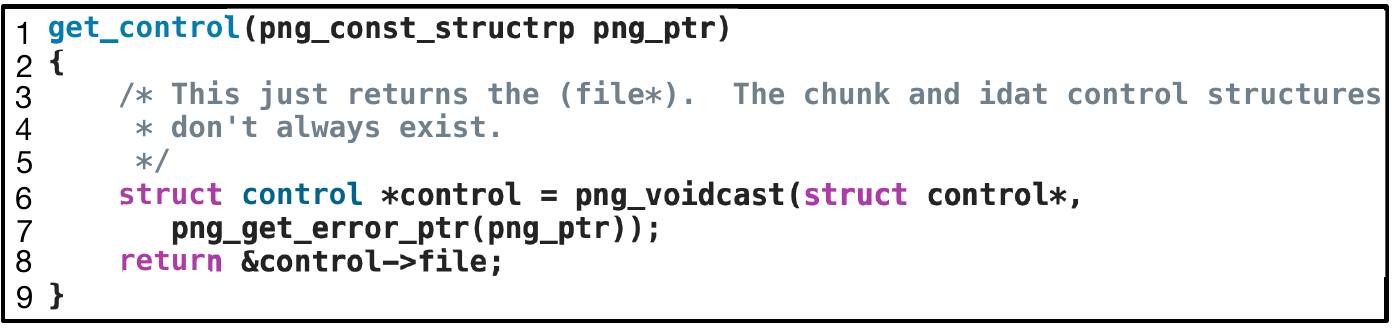}}
	\caption{A real-world vulnerability (i.e., CVE-2016-3751) in Android 4.x, allowing attackers to gain privileges via a crafted application, was not detected by our proposed method}
	\label{failed_case}
\end{figure}

As seen in the results of our previous experiment, the proposed StagedVulBERT achieves an overall F1 score of 92.26\% for coarse-grained vulnerability detection on the big\_vul dataset, which is the best performance achieved so far. That means our method can identify most vulnerabilities, including those undetectable by LineVul as shown in Fig.~\ref{VulExample}.
However, there are still some vulnerabilities our method fails to detect. For example, the ``get\_control" function shown in Fig.~\ref{failed_case} retrieves a control structure for PNG file handling, but it improperly casts pointers using png\_voidcast, leading to potential memory corruption or privilege gains. Thus, this function is a vulnerable example. However, our method fails to detect the vulnerability because it spans two functions (i.e., ``get\_control" and ``png\_voidcast"), which makes it difficult or even impossible to detect by analyzing each function individually. This limitation also exists in other transformer-based methods, which we will leave as future work by considering the dependencies among functions.

\section{Limitations and Future work}
\label{threats}
This paper proposes a novel detection framework StagedVulBERT and demonstrates its effectiveness when compared to DL-, transformer-, and LLM-based methods. However, it still has some limitations. First, while the big\_vul dataset is preferred by many baseline methods due to its real-world project origins and reflective sample imbalance, its labels, derived from bug-fix commits, might introduce noise that can impact model performance. Although this issue is recognized, as discussed in~\cite{croft2023data}, no ideal solution exists yet. Future work will focus on enhancing data quality research to minimize the influence of data noise on model training, as well as conducting more experiments on other datasets to evaluate the generalization ability of the proposed method.
Second, although our method performs well in detecting vulnerabilities related to C/C++, its effectiveness for other languages, such as Java, is yet to be determined. In future work, we plan to extend our approach to other programming languages. Since our approach does not rely on complex program parsing or compilation processes, it can theoretically be easily adapted to other programming languages.
Third, while our method can process longer code sequences through the proposed Algorithm 1, we currently truncate code samples when they exceed the token limit. This solution may not be optimal. Future work could explore using specific filtering heuristics to preserve more useful information and improve performance.
In addition, future research could benefit from incorporating contextual information such as AST, CFG, and PDG for enhanced vulnerability detection. Although it has been emphasized in many studies, most perform experiments on only a small fraction of functions in the big\_vul dataset, as numerous functions cannot be accurately parsed by existing tools such as Joern~\cite{ni2023distinguishing}. Therefore, combining this information with transformer architectures still presents challenges in real-world applications but also holds promise for more accurate, fine-grained detection.

\section{Related work}
\label{related-work}

With large amounts of code publicly available and the development of machine learning and DL technology, 
data-driven vulnerability discovery sheds new light on intelligent, automatic, and effective vulnerability detection. 
DL-based detection methods (e.g., VulDeePecker~\cite{li2016vulpecker} and SySeVR~\cite{li2018sysevr}) directly learn vulnerability patterns from the original source code, which can reduce the time and labor cost significantly due to no need for human-defined features. In addition, since many vulnerabilities can only be effectively detected by simultaneously considering the structure, control flow and dependencies within code, researchers have proposed several graph neural network-based models (e.g., Devign~\cite{zhou2019devign}, Vu1SPG~\cite{zheng2021vu1spg}, DeepWukong~\cite{cheng2021deepwukong}, GINN~\cite{wang2020learning}, Reveal~\cite{chakraborty2021deep} and AMPLE~\cite{wen2023vulnerability}) to learn a rich set of code semantics from these representations. They claim that their preliminary experiments show promising results,  but there remains potential for improvement in vulnerability detection of big functions with excessively long tokens~\cite{chakraborty2021deep}. 
In recent years, some transformer-based methods (e.g.,  LineVul~\cite{fu2022linevul}, VulBERTa~\cite{hanif2022vulBERTa} and SVulD~\cite{ni2023distinguishing}) have been proposed for vulnerability detection and have achieved great performance. This is supported by a review paper~\cite{steenhoek2023empirical} which compared 11 cutting-edge deep learning frameworks, highlighting LineVul's significant advantages on well-known datasets such as Devign and Big-Vul. LLMs have also been explored for this task, but their performances are weaker than those of specific transformer-based methods. Empirical studies~\cite{yin2024multitask} have confirmed the aforementioned findings.

In addition, the granularity of the aforementioned DL-based vulnerability detection methods is relatively coarse, primarily enabling detection at the function or code snippet levels. To achieve fine-grained vulnerability detection, some studies have proposed the following approaches: (1) interpreting prediction results using the attention mechanism employed in the coarse-grained detection model architecture~\cite{fu2022linevul}; (2) reusing statement representations extracted via GAT to provide statement-level detection~\cite{hin2022linevd}; (3) using gradients or hidden feature mapping values as approximations of input importance to identify tokens that may lead to vulnerabilities~\cite{baldassarre2019explainability, pope2019explainability}. However, these methods often struggle with generalization across different models. Therefore, some research has introduced the following interpretability methods, each specifically crafted to be compatible with a range of coarse-grained detection approaches. For instances,
Studies~\cite{funke2020hard, luo2020parameterized, schlichtkrull2020interpreting, wang2020causal, ying2019gnnexplainer,li2021vulnerability} explore how detection model outputs change when inputs are perturbed, thereby identifying inputs that significantly contribute to the predictions. Other research, such as~\cite{schnake2021higher, schwarzenberg2019layerwise}, involves decomposing a model's predictions into several components to evaluate the importance of input features. Additionally, there are approaches that train surrogate models, which mimic the behavior of the more complex original models, thus providing insights into the prediction mechanisms of these original detection models~\cite{huang2022graphlime, vu2020pgm, zhang2021relex}.

\section{Conclusion}
\label{conclusion}
In this paper, we introduce a PCL-based vulnerability detection framework targeted at both coarse- and fine-grained level detection.
We first leverage the proposed CodeBERT-HLS to automatically learn representations for code samples at different levels of semantic abstraction. Then, based on the learned representations, we develop a coarse-grained and a fine-grained vulnerability detection model based on DL. Benefiting from our novel PCL model and the pre-training task, our framework demonstrates significantly promising performance on a large-scale real-world dataset, as evidenced by our empirical experiments.
Notably, in fine-grained vulnerability detection, our method achieves a 75.17\% improvement in Top-5\% accuracy compared to the state-of-the-art method. 


%

\ifCLASSOPTIONcompsoc
  \section*{Acknowledgments}
\else

  \section*{Acknowledgment}
\fi

This work was supported by the National Natural Science Foundation of China (Grant Nos.62302125 and 62272132) and the Heilongjiang Postdoctoral Fund (No.LBH-Z23019).

\ifCLASSOPTIONcaptionsoff
  \newpage
\fi



\bibliographystyle{IEEEtran}
\bibliography{reference}

\begin{IEEEbiography}
	[{\includegraphics[width=1in,height=1.25in,clip,keepaspectratio]{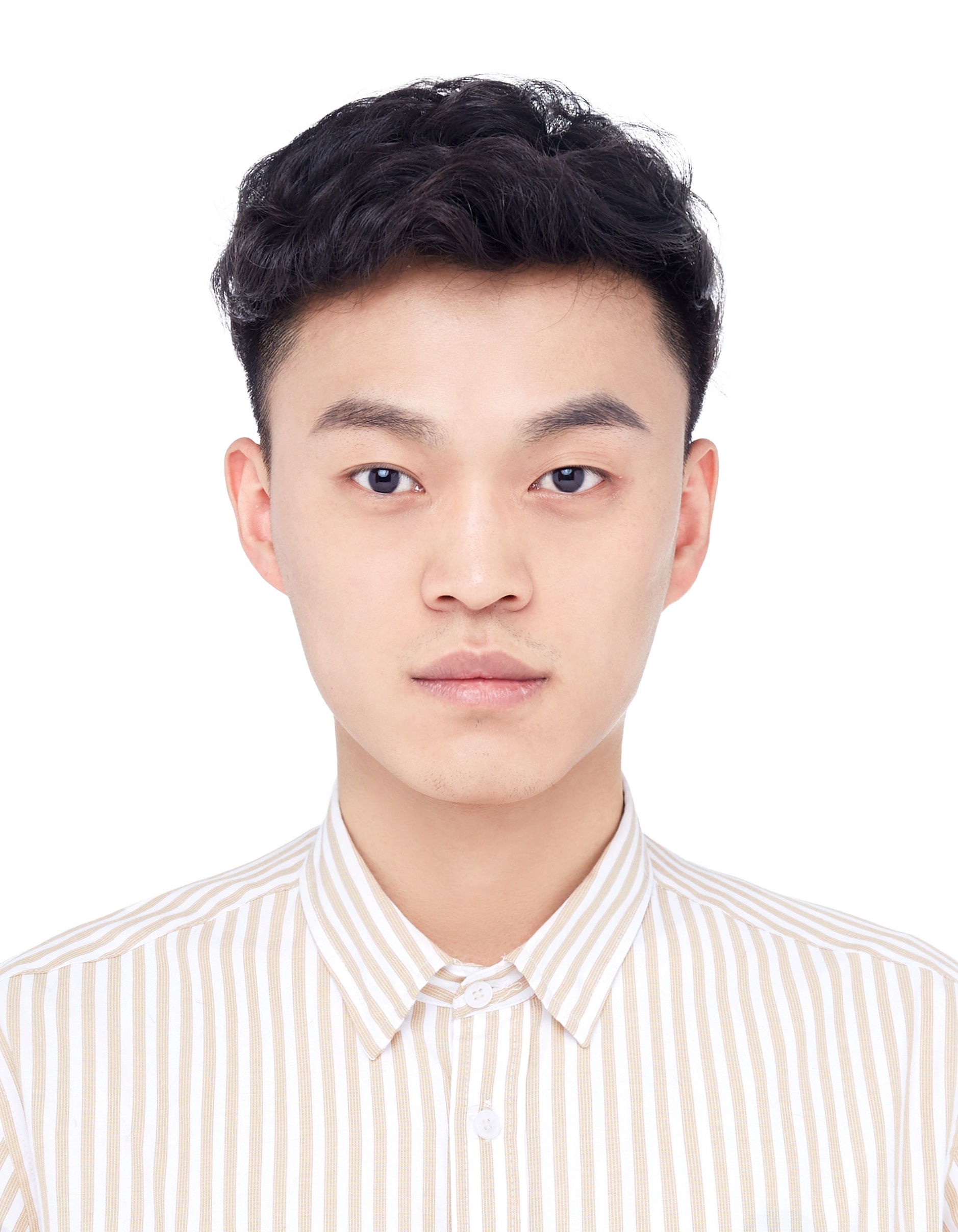}}]{Yuan Jiang}
	 is an assistant professor at the School of Computer Science and Technology, Harbin Institute of Technology. His main research interests include pre-trained code language models, mining software repositories, software vulnerability detection, and source code representation.
\end{IEEEbiography}

\begin{IEEEbiography}
	[{\includegraphics[width=1in,height=1.25in,clip,keepaspectratio]{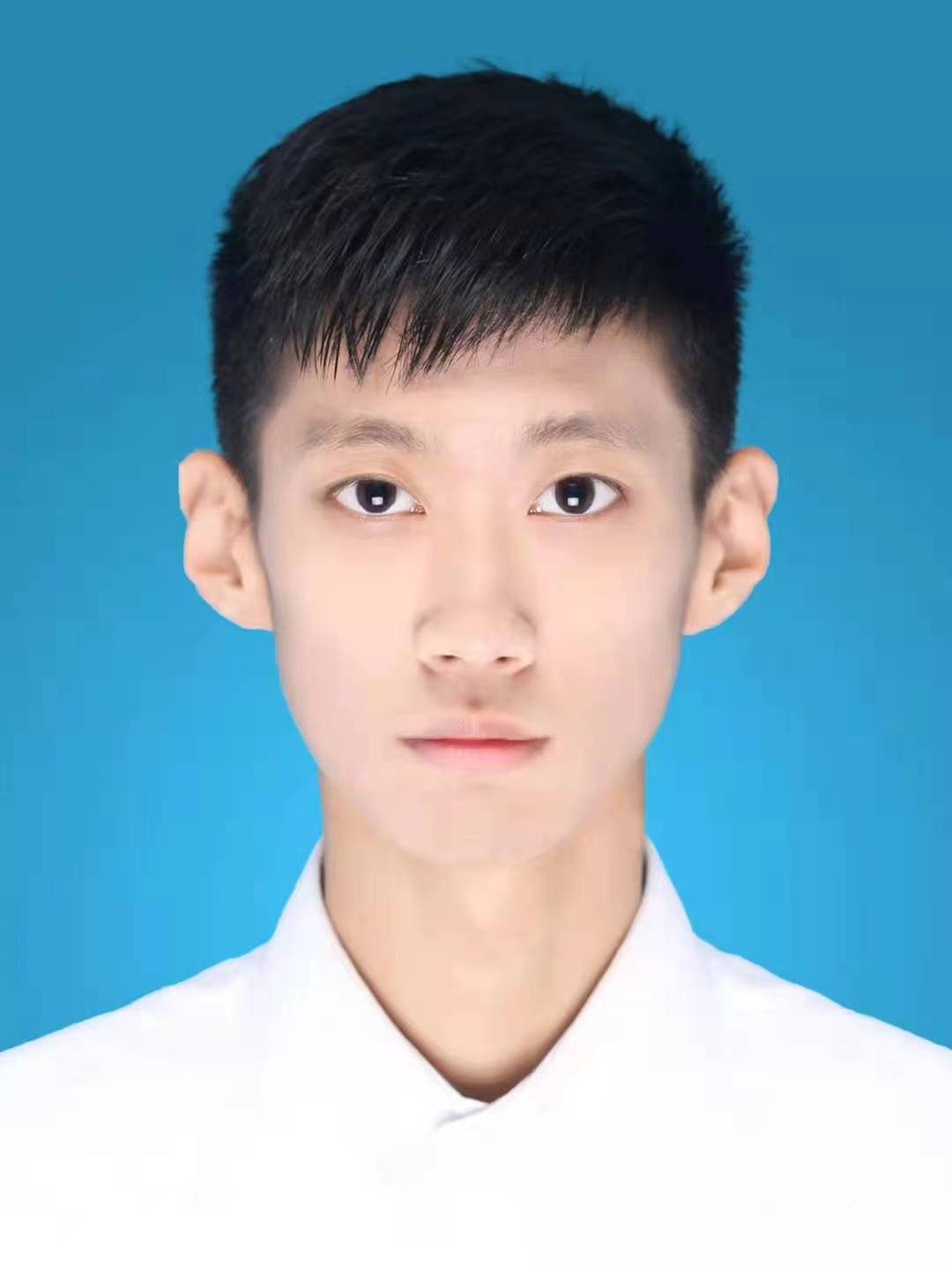}}]{Yujian Zhang}
	 is a graduate student at the School of Computer Science and Technology, Harbin Institute of Technology. His main research interests include software vulnerability detection and pre-trained code language models.
\end{IEEEbiography}

\begin{IEEEbiography}
	[{\includegraphics[width=1in,height=1.25in,clip,keepaspectratio]{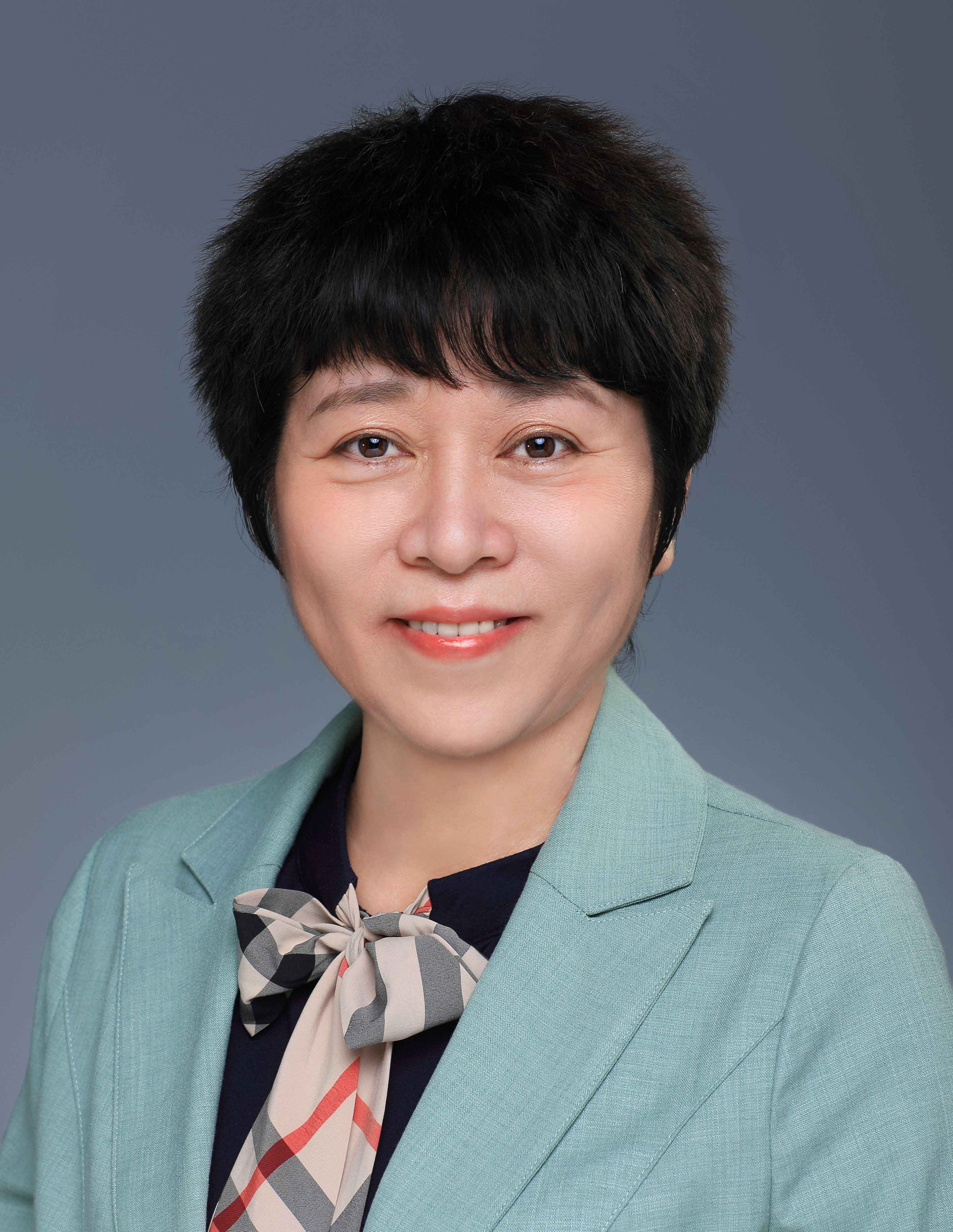}}]{Xiaohong Su}
	is a professor at the School of Computer Science and Technology, Harbin Institute of Technology. Her research interests include Intelligent software engineering, software vulnerability identification, code representation learning, bug triaging and localization, clone detection, and code search.
\end{IEEEbiography}

\begin{IEEEbiography}
	[{\includegraphics[width=1in,height=1.25in,clip,keepaspectratio]{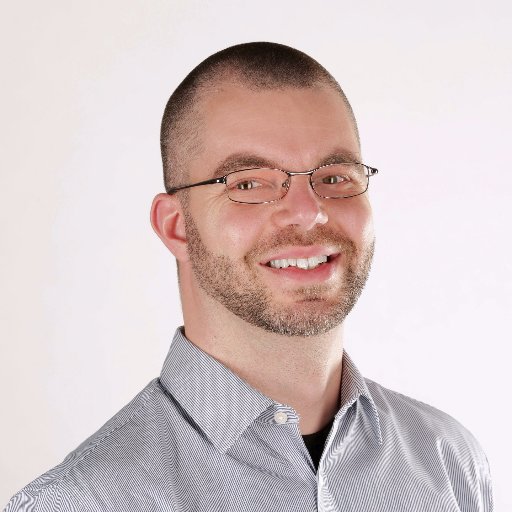}}]{Christoph Treude} is an Associate Professor of Computer Science at Singapore Management University. His notable achievements include receiving an ARC Discovery Early Career Research Award (2018-2020) and securing funding from industry giants such as Google and Facebook. Treude has been honored with four best paper awards, including two ACM SIGSOFT Distinguished Paper Awards. Currently, he serves on the Editorial Boards of the IEEE Transactions on Software Engineering and the Springer journal on Empirical Software Engineering. Additionally, he is the Open Science Editor for the Elsevier Journal of Systems and Software and has chaired conferences such as ICSME 2020, ICPC 2023, and TechDebt 2023. 

\end{IEEEbiography}

\begin{IEEEbiography}
	[{\includegraphics[width=1in,height=1.25in,clip,keepaspectratio]{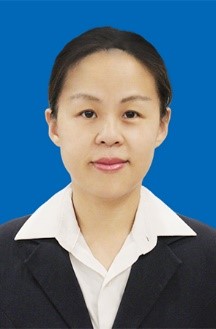}}]{Tiantian Wang}
	born in 1980. Received the Doctor’s degree from Harbin Institute of Technology, Harbin, Heilongjiang, China, in 2009. Since 2013, she has been an Associate Professor in computer science department of Harbin Institute of Technology. Her current research interests are software engineering, program analysis and computer aided education.
\end{IEEEbiography}




\end{document}